\documentclass[11pt]{article}
\usepackage[letterpaper,margin=1in]{geometry}
\usepackage[utf8]{inputenc}
\usepackage{graphicx}
\usepackage{natbib}
\usepackage{amsmath,amssymb}
\usepackage{booktabs}
\usepackage{threeparttable}
\usepackage[nodayofweek]{datetime}

\usepackage[
  pagebackref,
  colorlinks=true,
  linkcolor=blue,
  citecolor=blue,
  urlcolor=blue
]{hyperref}

\usepackage{tabularx}

\newcommand{\Normal}{\operatorname{Normal}}
\newcommand{\LS}{\operatorname{LocationScale}}
\newcommand{\Bernoulli}{\operatorname{Bernoulli}}
\newcommand{\POD}{\operatorname{POD}}
\newcommand{\E}{\operatorname{E}}

\newcommand{\footremember}[2]{%
  \footnote{#2}
  \newcounter{#1}
  \setcounter{#1}{\value{footnote}}%
}

\newif\ifempty

\newcommand{\Rsoftware}{\texttt{R}}

\newcommand{\norm}{\textrm{norm}}
\newcommand{\sev}{\textrm{sev}}
\newcommand{\lev}{\textrm{lev}}
\newcommand{\logis}{\textrm{logis}}

\newcommand{\muhat}{\hat{\mu}}
\newcommand{\sigmahat}{\hat{\sigma}}

\renewcommand*{\backrefalt}[4]{%
  \ifcase #1
  \or
    [#2]%
  \else
    [#2]%
  \fi
}

\newdateformat{mydate}{%
  \twodigit{\THEDAY}\ \monthname[\THEMONTH]\ \THEYEAR}

\graphicspath{{figures/}}
\title{Model-Assisted Probability of Detection Studies for Image-Based Human Inspection}

\author{Qizheng Xia
    \footremember{isu1}{
    Department of Industrial and Manufacturing Systems Engineering,
    Iowa State University,
    Ames, IA 50010, USA
    }
\and
William Q. Meeker
    \footremember{isu2}{
    Department of Statistics,
    Iowa State University,
    Ames, IA 50010, USA
    }
\and
Joseph N. Gray
    \footremember{cnde}{
   Center for Nondestructive Evaluation,
    Iowa State University,
    Ames, IA 50010, USA,
    }
\and
Qing Li
    \footremember{isu3}{
    Department of Industrial and Manufacturing Systems Engineering,
    Center for Nondestructive Evaluation,
    Iowa State University,
    Ames, IA 50010, USA,
    qlijane@iastate.edu
    (Corresponding author)
    }
}

\date{22 November 2026}

\begin{document}

\maketitle

\begin{abstract}
An important element in determining the reliability of components in
aging systems is the establishment of a probability of detection
(POD) curve for the inspection used. Historically, POD
curves have been estimated from a set of physical specimens
containing representative flaws and inspectors who record
detection results. Such experiments can be expensive, particularly
because of the cost of fabricating specimens with realistic flaws.

This paper demonstrates how
appropriately validated physics-based simulation can substantially
reduce, and in suitable applications potentially replace, costly
physical POD experiments.  Nondestructive evaluation (NDE)
simulation provides an alternative that can generate images
economically while representing a broader range of flaw
morphologies, locations, and other sources of inspection variability
than would generally be practical in a physical experiment. In this
study we use XRSIM, an X-ray radiographic simulator, to investigate
the POD for inspection of a weld containing flaws of varying size
and location. Flaw characteristics were randomly generated from
specified distributions, and flaw locations and the number of
flaws per image were also randomized, with images containing from
zero to three flaws.
Each individual simulated flaw is treated as an observational unit,
so an image containing multiple flaws contributes
multiple flaw-level hit-miss responses.

To include human inspection variability, the simulated images were
presented in random order to five inspectors, with each inspector
inspecting the images, twice. The resulting hit-miss data were
analyzed using a repeated-measures binary-regression model that
accounts for variability among flaws, inspectors,
flaw/inspector combinations,
and inspection-to-inspection. The fitted model was then used to
estimate the population-mean POD curve and corresponding statistical
uncertainty. Our results show that flaw-to-flaw variability
dominates, with appreciable inspector-to-inspector differences.
\end{abstract}


\newpage

\section{Introduction}
\label{sec:introduction}

\subsection{Background and Motivation}
\label{sec:background-and-motivation}
An experimental POD measurement is designed to sample the NDE signal
variability of a typical inspection. Sources of variability in NDE
inspection include
instrumentation variability, setup variability, flaw morphology
variability, human perception variability, data viewing variability
(particularly grayscale image viewing), and others. The cost of
preparing these experimental measures of POD is largely controlled
by a limited number of sample coupons capturing the inspection
geometry of both the part and the flaw morphology, by the limited
number of inspectors available in the trials and finally by the
range of equipment available for the study as compared to the
possible range of equipment that is typically used in the
field. Because of these limitations and doubts that an affordable
experimental POD study adequately samples the multi-parameter space,
alternatives such as NDE simulators become attractive.

Throughout this paper, we use the general term ``flaw'' for a
feature to be detected in an image. In our application, the
flaws are cracks, and we use the more specific term ``crack''
when discussing them.

\subsection{Challenges in Image-Based Human Inspection}
\label{sec:image-based-human-inspection}

\begin{figure}[htbp]
\centering
\includegraphics[width=0.7\linewidth]{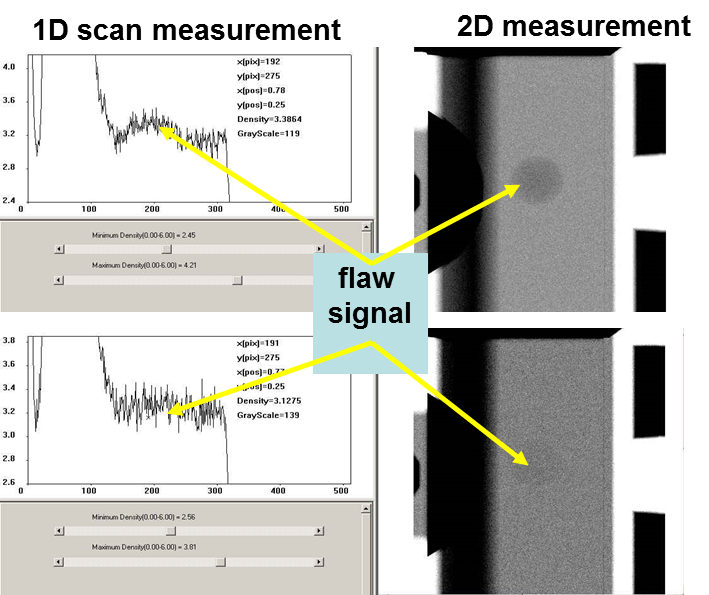}
\caption{The image at the top right has a CNR of 0.6:1 while the
  image at the bottom right has a CNR of 0.3:1. The plots to the
  left show a horizontal line trace across the center of the flaw
  and clearly show the low CNR.}
\label{fig:snr-examples}
\end{figure}

The example of using NDE simulators in this study focuses on
two-dimensional (2D) images typical of X-ray radiography and uses
the X-ray simulation program, XRSIM
\citep{xu1994,jensen1995,elshafiey1996,jensen1996,roth2011}. Most
NDE measurements are done with large probes, large in the sense that
the flaw typically detected is a small fraction of the area of the
probing energy. Examples of this include ultrasonic inspection,
where the beam size is generally larger than the reflector or in
eddy current inspection where the induced current flow by the coil
is large compared to the feature generating the indication. In this
type of measurement the signal size is roughly related to the flaw
size. In X-ray imaging the detector pixel size is often much smaller
than the size of the flaw of interest. In fact, if the flaw size is
smaller than a pixel, the flaw is not viewed as detectable. This is
true even for very high contrast features with contrast-to-noise
ratios (CNR) considerably larger than 3:1.  This observation about
2D imaging NDE methods where the pixel (point measurement) is small,
less than 150 microns, leads to an important difference in defining
what is detectable. In X-ray images, a CNR less than 1 is easily
detected. In Figure~\ref{fig:snr-examples} we see two flaws with
CNRs of 0.6 and 0.3, respectively, where the image provides leverage
in the detection limit. The role of detector resolution,
specifically pixel size, is important in any X-ray inspection.

In Figure~\ref{fig:snr-examples}, we also see on the left of each
image the line trace through the flaw showing the background noise
and the flaw contrast. In each image, seen on the right side, the
flaw is easily seen. The line traces give a clear indication of the
low CNR for the images. This multiple point measurement of a flaw by
several hundred pixels uses a property of the human perception
system’s ability to see patterns in noise. The likelihood that noise
caused all the pixels in the flaw image to have the same slight
change bias in the background mean value is very small. The
consequence of this is that the loss of detectability in X-ray
images occurs when the contrast is at the noise level and is not
governed by the large 3$\sigma$ deviations required by point
detectors. There are very few false calls due to noise fluctuations.

In generating a set of data for use in estimating POD we need to
capture the variability of the experimental measures of the flaw
signals and capture the variability of the human interpreting the
data. The factors that affect the flaw signal include flaw size and
morphology, the position of the flaw within the part, the geometry
of the part, and characteristics of the measurement system, such as
detector type, the X-ray source, and spatial resolution. Additional
parameters also come into play including the data
viewing conditions, the initial flaw distribution as controlled by
the manufacturing or aging process for the part, and the inspector
variability including the training of the inspector. Many
inspections based on 2D datasets cannot be reduced to one parameter
as is often used in, for example, when POD is modeled as a function
of crack length. The 2D nature of the data yields more dimensional
information than a single point measurement. Finally, some
inspections have a direct relationship between
performance degradation due to a single parameter for a flaw while
others have more complex relationships. In the case of a crack, both
length and depth through the material can affect structural
integrity, even though the probability of X-ray detection may be
governed by other flaw characteristics.

In using NDE simulation programs to generate synthetic typical NDE
data we have an opportunity to generate data that covers a range of
inspection variables that traditionally have not been included in
experimental POD studies due, mainly, to cost limitations. The best
example of this is the complexity of the part shape and especially
the complexity
of the flaw morphology. To fabricate flaws at any location with
morphologies typical of real parts is prohibitively costly. Even a
modest range of several hundred flaws in a number of samples is
costly. The limited usability of these samples is illustrated in
that should the alloy of the material change, a new study is often
required for that new material. If the loads on the component change,
generating different flaw populations, a new study and new samples
are required.

\subsection{Related Work}
\label{sec:related-work}
POD provides a statistical measure of an inspection procedure’s
capability to detect flaws. Traditional POD studies estimate the
probability that an inspection procedure detects a flaw as a
function of a flaw size or other relevant flaw parameter. The common
hit/miss and signal-response approaches are useful because they
reduce inspection performance to quantities such as $a_{90}$ (the
flaw size at which the estimated probability of detection is 0.90,
or equivalently, the probability of a miss is 0.10) and $a_{90/95}$
(the upper 95\% confidence bound for $a_{90}$), but they also
require enough flaw samples to represent the variability of the
inspection. Model-assisted POD (MAPOD) was developed to reduce this
experimental burden by combining empirical observations with
information from physics-based models.  \citet{li2014} used a
physical model for POD estimation in NDE and showed how model
information can be used to supplement limited experimental
data. \citet{legratiet2016mapod} presented a progressive MAPOD
methodology in which simulation, uncertainty quantification and POD
estimation are linked in one analysis process. \citet{bismut2021nde}
reviewed probabilistic NDE models and emphasized that POD should be
interpreted together with the inspection decision and the
uncertainty sources that affect the measurement. These studies
provide the general methodological basis for using simulation in a
POD study, but they do not address the specific image-formation and
human-viewing issues that arise in X-ray radiography.

Several studies have treated X-ray inspection capability by
modeling the X-ray process itself.
\citet{koshti2014xrayparameter} proposed an X-ray flaw size parameter
for POD studies that combines crack geometry and inspection setup
factors into a parameter intended to correlate monotonically with
X-ray image contrast and POD. That work argues that crack
detectability in radiography is not controlled by a single physical
dimension alone, and X-ray angle, detector geometry, crack opening,
part thickness and detector response can all change the effective
indication. In a later study, \citet{koshti2015contrast} used
simulated contrast and SNR calculations to help set up
radiographic techniques with desired flaw detectability, showing how
variations in crack depth, crack width, detector resolution, source
size and geometry change the predicted X-ray flaw parameter and
image contrast. These two models use rectangular slots in plates
and rely on response calibration, with scatter assumed to be low,
uniform, and controlled. \citet{koshti2017grazing} also extended this
modeling to grazing-angle X-rays and showed that low-energy X-rays
incident near the crack-face direction can change crack contrast
through specular reflection between smooth, flat crack faces. The
applicability of this effect is constrained by material attenuation
at low X-ray energies. However, these studies mainly define or
simulate X-ray flaw parameters and do not collect repeated human
hit/miss data from simulated radiographic images.

\subsection{Outline of the Proposed Study}
\label{sec:paper-organization}
We use a MAPOD framework for image-based human inspection in X-ray
radiography. The study uses XRSIM to generate synthetic weld-flaw
images with controlled variation in flaw size, flaw morphology,
flaw location, flaw orientation with respect to the incident
X-ray beam, and typical weld beam thickness. 
Unlike simulation-only
or automated image-analysis studies, the final inspection response
is obtained from human viewers under a controlled viewing protocol,
allowing inspector variability and viewing effects to be represented
in the POD analysis. The resulting hit-miss data are analyzed using
both a conventional single-inspection binary regression model and a
repeated-measures Bayesian random-effects model that accounts for
flaw-to-flaw, inspector-to-inspector, and repeated-inspection
variability. By combining physics-based image simulation with
controlled human inspection in this way, the study shows how
simulated radiographic images can substantially reduce the
experimental burden of POD studies without relying entirely on
costly physical flaw specimens.

\subsection{Overview}
\label{sec:paper-overview}
The remainder of this paper is organized as
follows. Section~\ref{sec:methods} describes the
simulated image generation, flaw parameter
selection, and inspection/viewing protocol.
Section~\ref{section:statistical.models.hit.miss.pod.data} describes
statistical models for hit-miss POD data and shows how to use them
to estimate POD.
Section~\ref{sec:experimental-study} describes the
inspection experiment, the resulting hit/miss
data, and the POD estimation results. Section~\ref{sec:discussion} discusses
implications for MAPOD studies and areas for future related research.

\section{Model-Assisted POD Methodology for Simulated X-ray
  Image Generation and Human Inspection Analysis}
\label{sec:methods}
\begin{figure}[htbp]
    \centering
    \includegraphics[width=\textwidth]{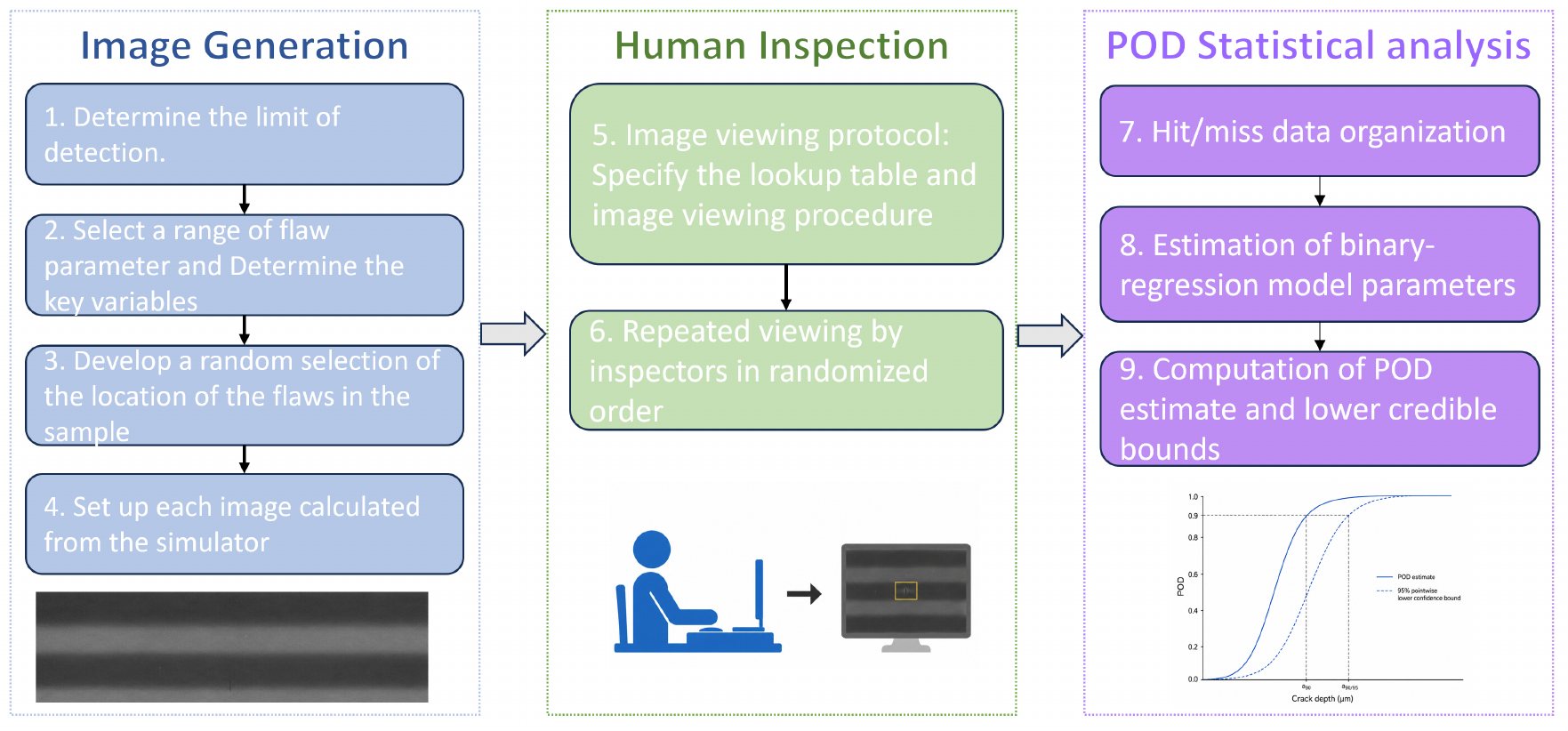}
    \caption{Workflow of the proposed model-assisted probability of
      detection study for image-based X-ray human inspection.}
    \label{fig:study-workflow}
\end{figure}

This section describes the methodology used to generate
the inspection data. XRSIM is used to generate
synthetic radiographic images of a weld
with realistic weld bead geometry and
flaws with controlled
variation in flaw characteristics, flaw locations, and
material-thickness conditions. This process includes selecting the
flaw type, defining the relevant flaw parameters, randomizing flaw
geometry and location, and specifying the image viewing protocol,
including grayscale mapping and inspection coverage.
Section~\ref{section:statistical.models.hit.miss.pod.data}
outlines appropriate statistical methods for
repeated-measures POD data and
Section~\ref{sec:repeated-measures-results} applies those methods
to estimate POD.
Together, these steps provide a model-assisted framework
for combining physics-based image simulation and human inspection
for POD estimation. The overall procedure of this study is shown in
Figure~\ref{fig:study-workflow}. The workflow first generates
simulated X-ray images with controlled flaw parameters and
locations, then collects repeated human hit/miss inspection
responses, and finally estimates POD curves using statistical POD
models.

\subsection{Simulation-based Image Generation}
\label{sec:simulation-based-image-generation}

\begin{figure}[htbp]
\centering
\includegraphics[width=0.92\linewidth]{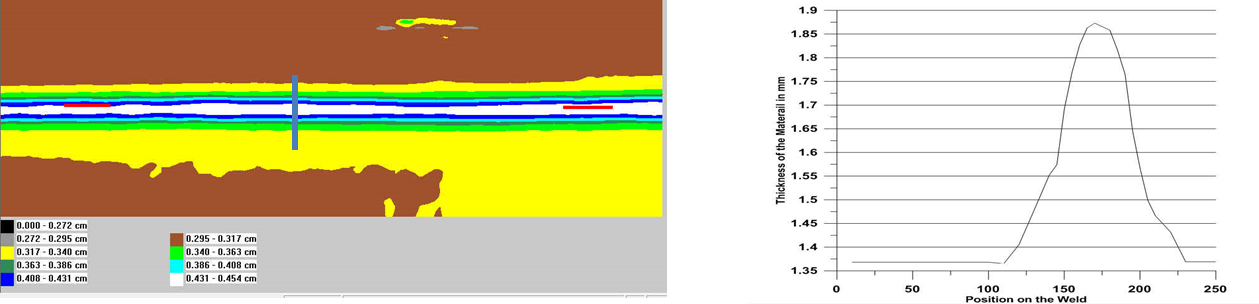}
\caption{The left image is a thickness map of a weld used in the
  flaw detectability study. The color scale represents the X-ray
  projection thickness. The plot on the right is the thickness at
  the location of the blue line on the map. The weld joins two
  plates. The two horizontal red lines are two flaws in different
  thicknesses of the weld material.}
\label{fig:weld-thickness-map}
\end{figure}

X-ray image generation for a POD study involves first determining
the limit of detectability for a range of material thicknesses
occurring in the part. This is done by using a step wedge with 32
steps with the maximum thickness and minimum step thickness
bracketing the range in the part.
For each step thickness a set of flaw sizes are placed ranging from
large, easily detected flaws to very small flaws below the contrast
detection limit. That detection limit is set by the noise in the
image formation process, see Figure~\ref{fig:snr-examples}.
This limit of detectability is
important due to the need to have a larger number of tests centered
on this flaw size with a sampling of flaws that are easy to see and
at the other end impossible to see. Using the computer-aided design
(CAD) model, a thickness map generated in XRSIM allows the thickness
range and,
importantly, the location of that variation, to be easily determined,
as illustrated in Figure~\ref{fig:weld-thickness-map}.

\subsection{Flaw Type and Flaw Parameter Design}
\label{sec:flaw-parameter-design}
The next issue that needs to be addressed is the range of flaws that
are in the initial flaw distribution. In a welding inspection, there
are several classes of flaws, including cracks, porosity, and
inclusions. The latter two are volumetric flaws. Cracks have strong
orientation dependence due to the very narrow crack width dimension
with respect to the two perpendicular dimensions. Crack flaws were
selected for this study due to their difficulty in detection using
an X-ray radiographic technique.  The resulting large number of
parameters influencing a detectable crack include the depth and
width of the crack which control the flaw signal. The crack
orientation to the incident X-ray beam is a very sensitive parameter
since for a very tight crack a few degree change in that orientation
causes the defect signal to change from an easily detected signal to
one with no contrast and, as such, undetectable. This is an
important point in that is shows that knowledge of the flaw
distribution is as important as the human factors variation of the
POD. A miss of a flaw due to the loss of the inspection signal is
not the same as an inspector’s temporary lapse in attention during
the inspection.

Crack opening is critical to X-ray inspection. Accordingly, only
surface-breaking, stress-free (open) cracks are considered, with
several crack-opening widths represented.
For the 105 synthetic flaws
that were generated, crack position in the image, position along the
weld, and location within regions of differing weld-bead thickness
were randomly selected. Thus, the simulated flaws were sampled from
an assumed population of cracks having varying orientation, depth, width, and
location within the weld bead both along the weld and in differing weld
bead thickness.

Due to the varying thickness of the part, two exposures are needed,
because for thin regions a proper exposure will leave the thick
regions underexposed. And for the second exposure the optimal
exposure for the thick region will saturate the thin regions of the
part.  The optimal condition for any exposure in a digital detector
is typically 80\% of the maximum bit depth. So for a 16-bit detector
that average grayscale would be 52,000 gray levels. The reason behind
this criterion is that the principal noise in an X-ray image is
controlled by total number of collected photons.  This process is
governed by a Poisson noise distribution in which the noise is
proportional to the square root of the total number of photons
collected by the detector.  The larger that photon count is, the
lower the Poisson noise. This noise controls the detector's contrast
sensitivity, specifically the minimum defect size detectable.

\subsection{Randomization of Flaw Characteristics and Locations}
\label{sec:randomization}
To generate the images, random numbers were computed for flaws
where the crack depth, width, location along the weld bead, and
location in the varying thickness of the weld were varied. The depth
was taken as the key parameter to reduce all the variables that
impact the flaw signal to one variable used in the hit/miss
and POD plots. The sensitivity plot for crack
depth against thickness was used to select more flaws in the range
of depth where the SNR was low. For each flaw,
the appropriate CAD model for the crack was scaled to match the flaw
parameters randomly selected from the initial distribution.
The CAD representation of the flaw was
then placed in the CAD model of the weld bead at the appropriate
location in the weld. This involves specifying the $(x,y,z)$
coordinates of the flaw's bounding box relative to the weld bead.

For comparison, manufacturing 100+ actual physical flaws at random
locations in a weld would be enormously expensive. Although creating
same number synthetic crack flaws using the CAD models and the XRSIM
interface is time-consuming, using simulated synthetic flaws is
tractable. It should be pointed out that using 3D computed
tomography to characterize a real crack with rough surface allows
the incorporation of a level of defect fidelity that is not possible
in fabricated flaw in test coupons.

\subsection{Image Viewing Protocol}
\label{sec:image-viewing-protocol}
\begin{figure}[htbp]
\centering
\includegraphics[width=0.7\linewidth]{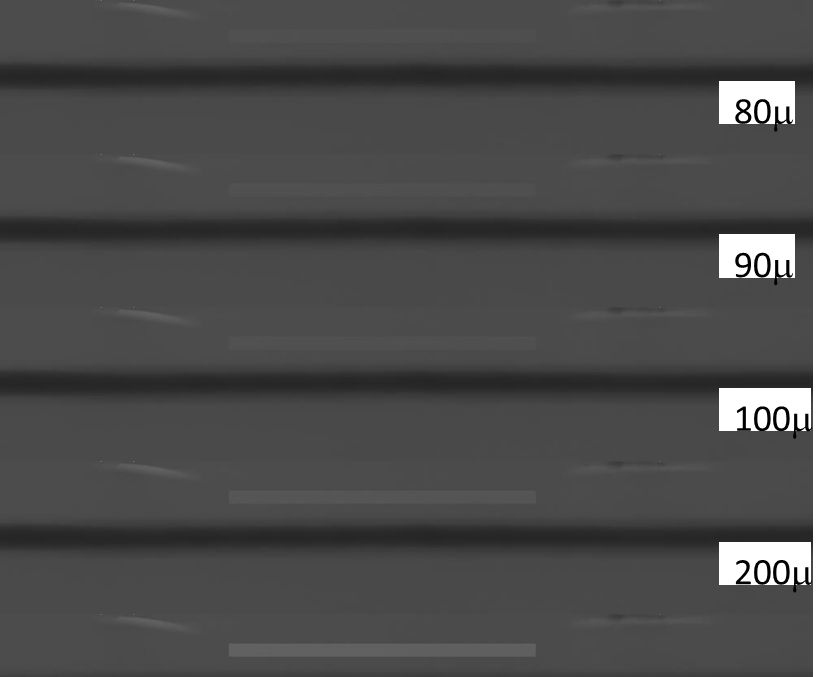}
\caption{Four images of a wide crack ranging from 80 microns to 200
  microns in depth are shown at identical locations. All of the
  parameters of the crack are larger than those used in the test
  images. The reduction in size illustrates the lower contrast of
  the flaw against the background. We note that these flaws are in
  the flat plate portion of the weld, a location
  where it is much easier to find
  a crack.}
\label{fig:wide-crack-depths}
\end{figure}

Examples of the images are shown in
Figure~\ref{fig:wide-crack-depths}. These examples were selected
from images containing larger flaws that are easily seen in printed
form due to the difficulty of adequately representing 14- or 16-bit
images in 8-bit images displayed by a computer monitor.
This limitation raises several issues that required
considerable care in establishing the inspection protocol as well as
training of the inspectors in viewing the images
\citep[e.g.,][]{KimpeTuytschaever2007,barten1992,barten1999}.

First, consider the issue of viewing 16-bit images. The noise level
in modern digital X-ray detectors is typically 50--70 gray levels
out of approximately 65,000 gray levels at the 16-bit saturation
level. Mapping the 16-bit grayscale to the 8-bit grayscale, which is
the limit of the monitor's display capability, is therefore a key
factor affecting inspector performance. In particular, an
appropriate mapping must be specified for the different thickness
ranges in the part because the grayscale changes as the weld-bead
thickness changes. With weld thicknesses spanning more than 6000
gray levels and flaw-detection sensitivity on the order of 70 gray
levels, the lookup table (LUT) used to examine the image must be
specified as part of the inspection protocol.

\begin{figure}[htbp]
\centering
\includegraphics[width=0.7\linewidth]{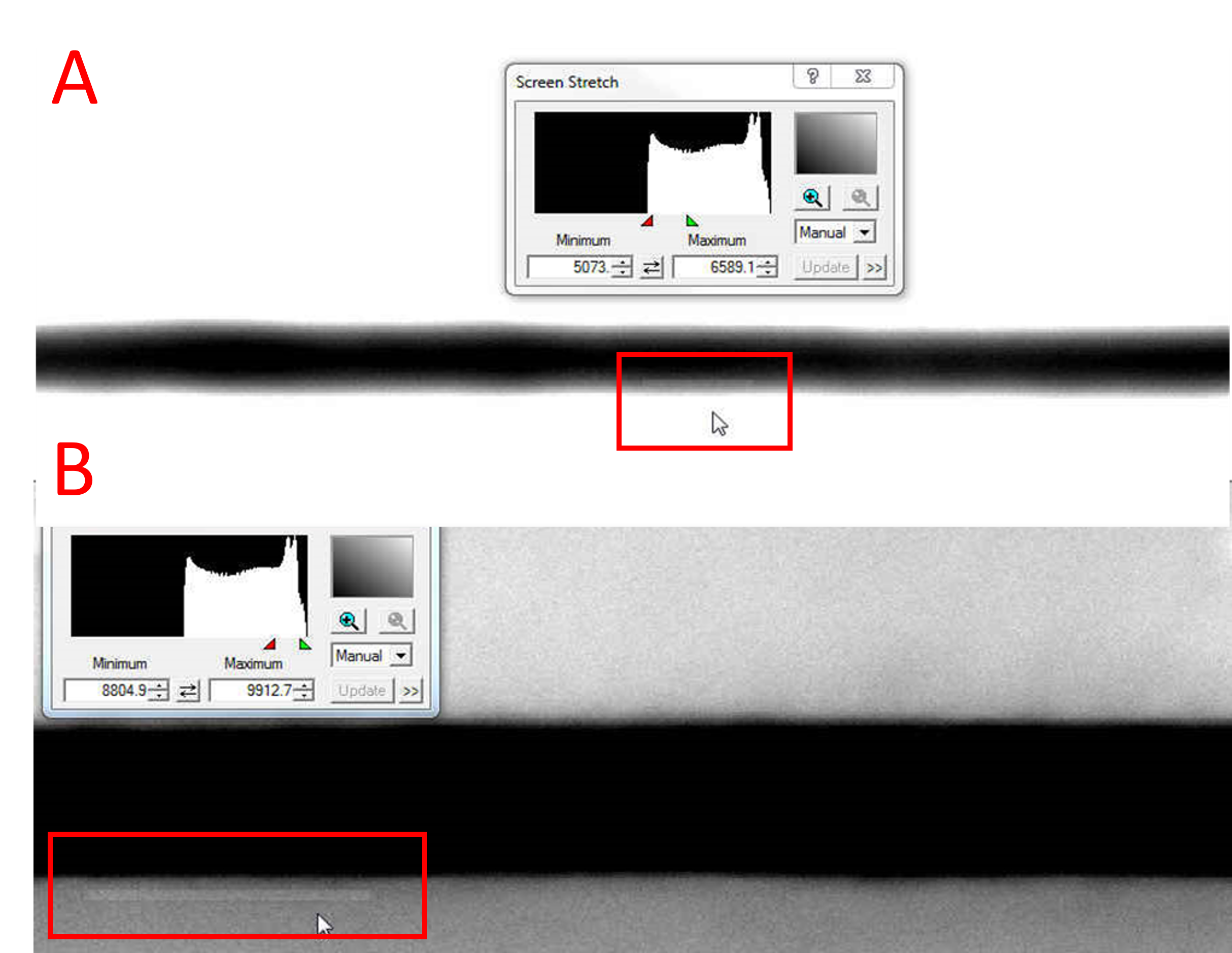}
\caption{Panels~A and B show two weld flaws at different locations
  in the weld bead that require different mappings from the
  sixteen-bit data to the eight-bit monitor.
  The small red and green triangles show the range of 16-bit
  data mapped to the 8-bit monitor gray scale.
  The cursor in each
  panel points to a crack flaw.}
\label{fig:grayscale-mapping}
\end{figure}

In Figure~\ref{fig:grayscale-mapping}, Panel~A has a 16-bit to
8-bit grayscale mapping (defined by a specific LUT) set for the
thicker part of the bead and has a very narrow grayscale range for
suitable viewing. Panel~B is more forgiving on the
grayscale range but also shows the impact of the LUT. In both panels
the LUT suitable for one thickness eliminates the possibility to see
the other, hence the need for multiple maps to view the image. Note
that the cursor points to the crack flaw in each panel.

A second issue in viewing an image is ensuring that the entire image
is systematically examined. Studies have shown that observers tend
to move their gaze among distinctive features in an image rather
than examining all regions uniformly
\citep[e.g.,][]{livingstone2008}. Eye-tracking studies, in which
cameras record an observer's gaze while viewing an image, demonstrate
this tendency. This issue is of particular concern in medical
imaging, where radiologists often examine large images. One approach
for encouraging complete coverage is to use a moving mask that
systematically rasters over smaller regions of the image, requiring
the observer to examine the image region by region. We developed a
simple mask for this purpose.

\subsection{Summary of Steps to Generate Images}
\label{sec:summary-of-steps}
To summarize, the steps needed to generate images for an inspector
hit-miss evaluation are:
\begin{itemize}

\item Determine the limit of detection using the simulation tool and
the orientation of the flaw relative to the X-ray projection direction.
This must be
  done over the range of thickness variation in the sample.

\item Select a range of flaw parameters representative of the
  initial flaw distribution expected in the inspection, and center
  the sampling of those flaws around the limit of detection. In this
  study, we selected a notch representation of a crack typical of
  those that can be manufactured for experimental determinations of
  POD.

\item Determine the key variables that influence flaw signals, in
  this case the crack opening, the tube voltage, and the location
  of the crack in the weld bead.

\item Randomly select the locations of the flaws in the sample. In
  this case, we varied location both across and along the weld bead.
  The number of flaws in an image ranged from zero to three.

\item Generate each image using the simulator, keeping a record of
  the actual locations of the flaws.

\item Develop an appropriate viewing protocol for the images,
  specifying the LUT used to map the 16-bit detector data to the
  8-bit computer display and developing a rastering mask to ensure
  that the entire image is viewed.

\item Develop a means of tracking the bookkeeping associated with
  randomization of the image order for inspection. We used file
  names with all key parameters embedded in them and created a
  mapping to a second set of file names containing numbers that
  specified the viewing order. We regenerated the viewing order for
  each inspection of the images.

\item Record the inspection results as hit-miss data, indicating
  which flaws present in each image were found and which were missed.

\end{itemize}

\section{Statistical Models for Hit-Miss POD Data}
\label{section:statistical.models.hit.miss.pod.data}

\subsection{Statistical Models for Single-Inspection  Hit-Miss POD Data}
\label{section:single-inspection-pod}
In a POD study where the available flaws (synthetic flaws in images for
our example) are inspected only once, the
simple binary regression model is often used
to analyze the data. A detailed description of this model is given
in \citet[][Appendix~G]{milhdbk1823a2009} and will
not be repeated here. We review only the basic ideas. The binary
random variable,
\begin{equation*}
H = \begin{cases}
  1, & \text{if the flaw is detected (a hit)},\\
  0, &  \text{if the flaw is not detected (a miss)}
\end{cases}
\end{equation*}
is the response. We let $x$ denote an appropriate (possibly
transformed) characteristic of the
flaw (e.g., crack length or flaw area on an image)
that has an important effect on POD. Then the binary regression model
for POD is
\begin{equation}
\label{equation:binary.glm}
\POD(x) = \Pr(H=1 | x) =\Phi(\beta_{0}+\beta_{1} x),
\end{equation}
where $\Phi(\cdot)$ is the standard cdf of a location-scale (LS) distribution. 
Commonly used LS distributions include the normal, logistic,
smallest extreme value, and largest extreme value distributions. The choice
of distribution corresponds to the choice of a link function (probit, logit,
complementary log-log, and log-log, respectively) in the generalized
linear model (GLM) formulation of binary regression. The standard cdfs for
these LS distributions are denoted by $\Phi_{\norm}$,
$\Phi_{\logis}$, $\Phi_{\sev}$, and $\Phi_{\lev}$. \citet[][Chapter~4 and
Appendix~B.3]{MeekerEscobarPascual2022} provide details
about these and other LS distributions.

It is useful to reparameterize the model in \eqref{equation:binary.glm} by using
\begin{equation}
\label{equation:binary.ls}
\POD(x) = \Pr(H=1 | x) =\Phi\left( \frac{x-\mu}{\sigma} \right).
\end{equation}
For $\beta_{1}>0$, which implies an increasing POD curve,
$\mu=-\beta_{0} / \beta_{1}$ and $\sigma=1 / \beta_{1}$. The important
advantage of this reparameterization is that $\mu$ and $\sigma$ have
the same units as flaw size $x$ and have useful interpretations. In
particular, $\mu$ is the location parameter of the LS distribution
describing the (random) flaw size that would just be detected on a
particular inspection and $\sigma$ is the corresponding scale
parameter describing how rapidly the POD curve increases with $x$.
For symmetric distributions (e.g., normal and logistic), $\mu$ is
the mean/median, and for the normal distribution, $\sigma$ is the standard
deviation. Another advantage of the $(\mu,\sigma)$ parameterization
is that these parameters are stable parameters and, unlike
$(\beta_{0},\beta_{1})$, are typically more readily identified and less
highly correlated.

Today, the most commonly used method of statistical estimation for
binary data is
maximum likelihood (ML). The method of ML finds those values of the
parameters that maximize the probability of the data. Details about
the method of ML, applied to estimation of POD, are given in
\citet[][Appendix~G]{milhdbk1823a2009} and will not be repeated
here. Statistical uncertainty (i.e., the uncertainty due to limited
data) is typically quantified by reporting standard errors of
estimates or confidence intervals.
\citet[][Appendix~G]{milhdbk1823a2009} also
describes methods for doing the related computations.

Traditionally, a set of pointwise lower confidence bounds
(sometimes called a confidence band) has been
reported, along with the point estimates of the POD function.
There are several methods
for computing confidence bounds, again described in
\citet[][Appendix~G]{milhdbk1823a2009}.
Although confidence interval methods that are based
on likelihood are known to have better theoretical properties, the
easy-to-compute ``Wald'' method is most commonly used and usually
provides an adequate approximation for sample sizes that are
typically used in NDE applications (e.g., 60 or more flaws in a
study using hit/miss data), provided that the data are not sparse in important
parts of the flaw-size range.

\subsection{Statistical Models for Repeated-Measures  Hit-Miss POD Data}
\label{section:repeated-measures-pod}
\subsubsection{Background}
An important limitation of the relatively simple statistical model
outlined in Section~\ref{section:single-inspection-pod} is that it is
not general enough to handle experiments where flaws are inspected
more than once. For example, if 100 flaws are available for
inspection and if all of the flaws are inspected by each of four
different inspectors, the amount of information available is not the
same as if one inspector inspected 400 different flaws. Special
statistical methods are required to estimate POD from
repeated-measures experiments. Such methods were previously
described for $\hat{a}$ versus $a$ POD data in~\citet{li2014}
and hit-miss POD data in~\citet{koh2014,KohMeeker2019}.

To properly analyze the repeated-measures POD data, it is necessary
to include in the model random effects that describe the other
sources of variability in the experimental data, such as inspector
differences, differences in flaw morphology, and differences in the
experimental setup.

\subsubsection{Experimental data structure and notation}
\begin{figure}[htbp]
\centering
\includegraphics[width=0.7\linewidth]{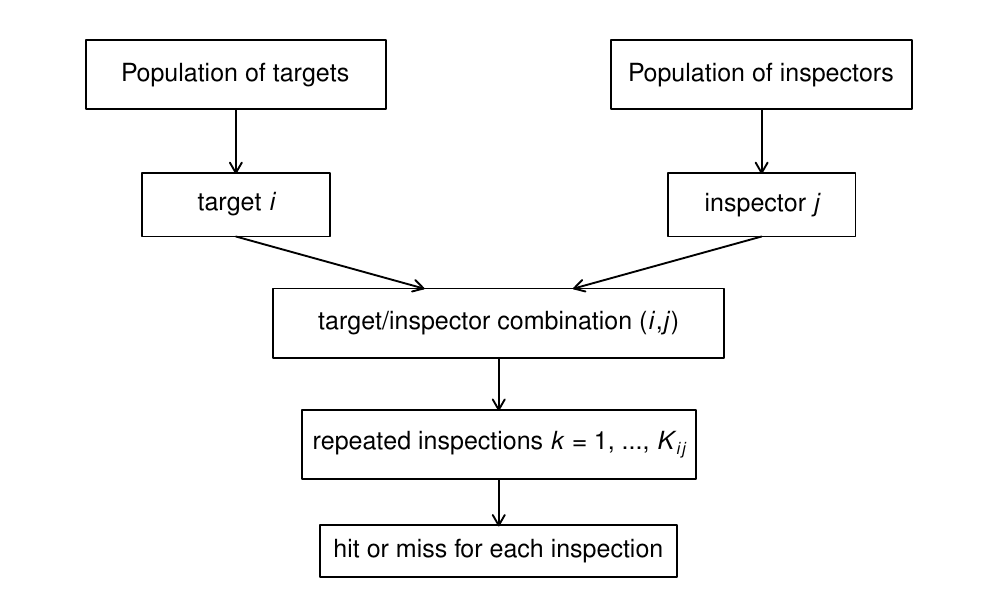}
\caption{Illustration of the experimental data structure, with flaws
  and inspectors crossed and repeated inspections nested within
  flaw/inspector combinations.}
\label{figure:Xray.hierarchical.data}
\end{figure}

Consider hit-miss POD data from an experiment in which flaws and
inspectors are crossed. Flaw $i$ has known size
$x_{i}$, $i=1, \dots, I$, inspector $j$, $j=1, \dots, J$ inspects the flaws,
and the same flaw/inspector combination can be inspected
repeatedly. Let $k$, $k=1, \dots, K_{ij}$ index the repeats nested
within flaw/inspector combination $(i,j)$. In a balanced
experiment, each inspector inspects all of the flaws and $K_{ij}$
is the same for all combinations of $i$ and $j$.
Figure~\ref{figure:Xray.hierarchical.data} illustrates the
experimental design and the resulting data.

\subsubsection{A latent random variable model formulation}
To aid in the explanation of the binary regression model, we
introduce an unobservable (sometimes called latent) continuous
random variable that can be interpreted as the flaw size that would
just be detected on a particular inspection:
\begin{equation}
Y_{ijk} = \mu  + \tau_{i} + \delta_{j} + \gamma_{ij} + \epsilon_{ijk},
\label{equation:threshold.model}
\end{equation}
where\par
\begin{tabularx}{\textwidth}
  {@{}l@{\quad}r@{\quad \quad}>{\raggedright\arraybackslash}X@{}}
  & $\tau_{i} \sim \Normal(0,\sigma_{\tau})$
  & is a random flaw-to-flaw effect due, for example, to random crack
    morphology. \\[0.5ex]
  & $\delta_{j} \sim \Normal(0,\sigma_{\delta})$
  & is a random inspector-to-inspector effect due, for example, to experience
    and skill. \\[0.5ex]
  & $\gamma_{ij} \sim \Normal(0,\sigma_{\gamma})$
  & is a random flaw/inspector-combination effect arising, for example, because
    an inspector $j$ may have difficulty with flaw $i$.\\[0.5ex]
  & $\mu$
    & is the location parameter for the distribution of
    $Y_{ijk}$ when $\tau_{i}=\delta_{j}=\gamma_{ij}=0$. \\
  & $\epsilon_{ijk} \sim \LS(0, \sigma)$
    & is a random error term having an LS distribution that describes
    additional inspection-to-inspection variability
    in the repeat inspections of flaw $i$ by
inspector $j$.
\end{tabularx}

\begin{figure}[htbp]
\centering
\includegraphics[width=0.7\linewidth]{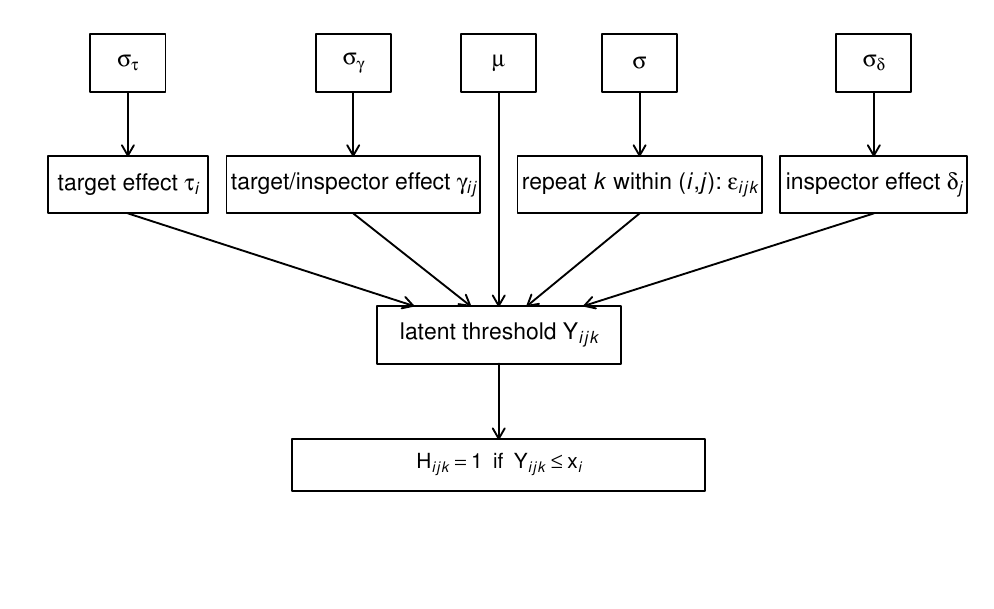}\\[-7ex]
\caption{Illustration of the hierarchical statistical model for the
  latent threshold $Y_{ijk}$ and its sources of variability.}
\label{figure:Xray.hierarchical.model}
\end{figure}

Figure~\ref{figure:Xray.hierarchical.model} illustrates the
hierarchical structure of the statistical model for $Y_{ijk}$.
From~\eqref{equation:threshold.model}, we see that an increase in any
of the random effects increases $Y_{ijk}$, which, in turn, implies a flaw
that is more difficult to detect.

Repeated inspections of the same flaw by the same inspector have
the same values of $\tau_{i}$, $\delta_{j}$, and $\gamma_{ij}$. Also, these
random effects and $\epsilon_{ijk}$ are mutually independent across
all values of their indices.
 The random-effect standard deviations
$\sigma_\tau$, $\sigma_\delta$, and $\sigma_\gamma$ describe
the flaw, inspector, and flaw/inspector-combination
variability, respectively.
The random variable
$\epsilon_{ijk}/\sigma$ has a standard LS distribution; thus,
\begin{equation}
\Pr\left(\frac{\epsilon_{ijk}}{\sigma}  \le z \right) = \Phi(z).
\label{equation:epsilon.cdf}
\end{equation}
Note that $Y_{ijk}$ and all five population-level parameters $\mu$, $\sigma$,
$\sigma_\tau$, $\sigma_\delta$, and $\sigma_\gamma$ have the same
physical units as flaw size. This provides an
important advantage over the
usual GLM binary regression formulation.

Recall that the number of flaws in an image ranged from zero to
three.  The hit-miss responses from multiple flaws in the same image
could be correlated because they have the same image-level
conditions.  However, because the simulated image backgrounds and
viewing conditions were carefully controlled, we expect such
dependence to be small relative to the substantial flaw-to-flaw
variability and therefore do not explicitly model image-level
effects.

\subsubsection{Conditional probability of a hit}
\label{section:conditional.distribution.probability.hit}
According to the definition of the latent variable $Y_{ijk}$, a hit occurs if
$Y_{ijk} \le x_{i}$. Using
\eqref{equation:threshold.model} and the result in
\eqref{equation:epsilon.cdf}, conditional on the fixed values of the
random effects $\tau_{i}$, $\delta_{j}$, and $\gamma_{ij}$, the
probability of a hit is
\begin{align}
p_{ij} =  \Pr(H_{ijk}=1 \mid \tau_{i},\delta_{j},\gamma_{ij}) &=
\Pr\left(
Y_{ijk}\le x_{i} \mid \tau_{i},\delta_{j},\gamma_{ij}
\right) \notag\\
&=\Pr\left[
  \epsilon_{ijk} \le x_{i}-(\mu+\tau_{i}+\delta_{j}+\gamma_{ij})
\mid \tau_{i},\delta_{j},\gamma_{ij}
\right] \notag\\
&=
\Phi\left[
\frac{x_{i} - (\mu+\tau_{i}+\delta_{j}+\gamma_{ij})}{\sigma}
\right].
\label{equation:conditional.hit.probability}
\end{align}
Note that $p_{ij}$ does not depend on $k$, so the $K_{ij}$ repeat
inspections of flaw $i$ by inspector $j$ have a common conditional
probability of a hit.

\subsubsection{Conditional Bernoulli likelihood}
\label{section:conditional.bernoulli.likelihood}
Modern Bayesian software (which we use for estimation in this paper)
requires specification of the likelihood of
the data, conditional on given values of any random effects in the
hierarchical statistical model.
Conditional on the fixed values of the random effects,
\begin{equation}
H_{ijk}\mid\tau_{i},\delta_{j},\gamma_{ij}
\sim
\Bernoulli(p_{ij}),
\label{equation:bernoulli}
\end{equation}
where $p_{ij}$ is given by \eqref{equation:conditional.hit.probability}.
Then the conditional likelihood is
\begin{equation*}
L_c(\mu,\sigma,\boldsymbol{\tau},\boldsymbol{\delta},\boldsymbol{\gamma})
=
\prod_{i=1}^{I}\prod_{j=1}^{J}\prod_{k=1}^{K_{ij}}
p_{ij}^{h_{ijk}}(1-p_{ij})^{1-h_{ijk}},
\end{equation*}
where $h_{ijk}$ is the realized 0/1 value of $H_{ijk}$ for
inspection $k$ by inspector $j$ on flaw $i$ and
$\boldsymbol{\tau}$, $\boldsymbol{\delta}$, and
$\boldsymbol{\gamma}$ are vectors that contain the random effects
over all values of their indices. Then \eqref{equation:bernoulli},
along with the normal distributions for $\tau_{i}$, $\delta_{j}$, and
$\gamma_{ij}$, specifies the hierarchical model. For Bayesian
inference, priors for the population-level parameters are also
required. The random effects $\tau_{i}$,
$\delta_{j}$, and $\gamma_{ij}$ for $i=1,\dots, I$ and $j=1, \dots, J$
are then estimated jointly with the population-level parameters $\mu$,
$\sigma$, $\sigma_\tau$, $\sigma_\delta$, and $\sigma_\gamma$.

\subsubsection{Conditional POD}
\label{section:conditional.pod}
For a flaw having size $x$, flaw effect $\tau$, inspector
effect $\delta$, and flaw/inspector-combination effect $\gamma$,
the conditional POD for one inspection is
\begin{equation*}
  \POD(x\mid\tau,\delta,\gamma)
  = \Pr(Y \le x\mid\tau,\delta,\gamma)
  =\Phi\left[
    \frac{x - (\mu+\tau+\delta+\gamma)}{\sigma}
    \right].
\end{equation*}
Note that $\mu+\tau+\delta+\gamma$ is the location of the
conditional distribution of $Y$, the flaw size that would just be
detected. For symmetric distributions like the normal and logistic,
this quantity is also the $a_{50}$ (median) value. As in
\eqref{equation:binary.ls}, $\sigma$ is the scale parameter and
determines how fast the POD curve increases with $x$.

\subsubsection{Population-mean (unconditional) POD}
\label{section:unconditional.POD}
For the population-mean POD, consider a flaw and an inspector selected
at random from their respective populations and a single inspection
of that flaw by that inspector. The effects $\tau$, $\delta$, and
$\gamma$ are then random, and the conditional POD must be averaged
over their distributions. Let $\psi=\tau+\delta+\gamma$. Then
$\psi \sim \Normal(0,\sigma_\psi)$, where
$\sigma_\psi=\sqrt{\sigma_\tau^2+\sigma_\delta^2+\sigma_\gamma^2}$, and
\begin{equation*}
\POD(x) =
\E\left[
\Phi\left(
\frac{x-(\mu+\psi)}{\sigma}
\right)
\right].
\end{equation*}
Using
\begin{equation*}
\psi=\sigma_\psi Z,
\qquad
Z \sim \Normal(0,1),
\end{equation*}
the expectation is evaluated with
\begin{equation}
\POD(x)  =
\int_{-\infty}^{\infty}
\Phi\left[
\frac{x-(\mu + \sigma_\psi z)}{\sigma}
\right]
\phi_{\norm}(z)\,dz.
\label{equation:unconditional.pod.integral}
\end{equation}

When the random error term $\epsilon_{ijk} \sim \Normal(0, \sigma)$
instead of the more general LS distribution, a simplification is available.
For a randomly selected flaw and inspector, the latent threshold is
\begin{equation*}
Y=\mu+\psi+\epsilon.
\end{equation*}
Because $\psi$ and $\epsilon$ are independent normal random
variables, the latent flaw size that would
just be detected is
\begin{equation*}
Y \sim
\Normal\left(
\mu,
\sqrt{
\sigma_\tau^2+\sigma_\delta^2+\sigma_\gamma^2+\sigma^2
}
\right).
\end{equation*}
Therefore
\begin{equation}
\POD(x) = \Pr(Y \le x)=
\Phi_{\norm}\left[
\frac{x-\mu}
{\sqrt{\sigma_\tau^2+\sigma_\delta^2+\sigma_\gamma^2 + \sigma^2}}
\right].
\label{equation:probit.unconditional.pod}
\end{equation}

\subsubsection{Conditional POD for a particular inspector}
\label{section:conditional.for.inspector.POD}
For a particular inspector $j$, the inspector effect $\delta_{j}$ is
fixed, while the flaw and flaw/inspector-combination effects are
random. Because
\begin{equation*}
\tau+\gamma \sim
\Normal\left(0,\sqrt{\sigma_{\tau}^{2}+\sigma_{\gamma}^{2}}\right),
\end{equation*}
the POD for inspector $j$, averaged over the population of flaws, is
\begin{equation*}
\POD_{j}(x)=
\E\left[
\Phi\left(
\frac{x-(\mu+\delta_{j}+\tau+\gamma)}{\sigma}
\right)
\right].
\end{equation*}
The expectation can be evaluated numerically as in
Section~\ref{section:unconditional.POD}. When
$\epsilon_{ijk}\sim\Normal(0,\sigma)$, the result simplifies to
\begin{equation}
\label{equation:normal.conditional.inspector.pod}
\POD_{j}(x)=
\Phi_{\norm}\left[
\frac{x-(\mu+\delta_{j})}
{\sqrt{\sigma_{\tau}^{2}+\sigma_{\gamma}^{2}+\sigma^{2}}}
\right].
\end{equation}
Thus a positive inspector effect shifts the POD curve toward larger
flaw sizes, whereas a negative inspector effect shifts it toward
smaller flaw sizes.

\subsubsection{Statistical estimation for the hierarchical POD model}
\label{section:statistical.estimation.pod.model}

Although it is possible to use ML methods to estimate the parameters
of the random effects model presented in this section, the
likelihood function requires evaluating complicated high-dimensional
integrals over the random effects. \Rsoftware{} packages such as
lme4 and glmmTMB avoid direct numerical integration by using
approximations to the likelihood. Computing estimates and
approximate confidence bounds for \textit{functions} of the
parameters (like POD) adds another layer of complexity, as described
in Section~\ref{section:unconditional.POD}. Also, because the number
of inspectors and repeat inspections is small in our experiment, the
``large-sample'' approximate methods for computing confidence
intervals may not provide accurate approximations to
the desired level of confidence.

A useful alternative to ML in these situations is Bayesian inference
with weakly informative prior distributions, employing Markov chain
Monte Carlo (MCMC) methods to compute draws from the needed
posterior distributions. This approach avoids the likelihood
approximations required by ML software, and posterior draws for
functions of the parameters (like POD) are readily obtained by
evaluating the desired function at each posterior draw. Credible
intervals constructed from these draws do not rely on large-sample
normal-approximation theory. Such methods are readily accessible
using modern software such as Stan
\citep[][]{Stan-software}. Bayesian methods have also been used for
similar POD applications by \citet{li2014} and
\citet{koh2014,KohMeeker2019}.

Modern statistical software for Bayesian analysis allows
specification of the statistical model (including prior
distributions) using an intuitive modeling language. Some care is
needed in parameterizing the model and choosing appropriate prior
distributions. In many applications, weakly informative prior
distributions can be chosen so that they stabilize estimation of
weakly identified parameters while allowing the information in the
data to dominate. With appropriately chosen priors, Bayesian
credible bounds can also have good frequentist coverage properties
(e.g., a nominal 95\% lower credible bound can have approximately
95\% repeated-sampling coverage). These issues are discussed and
illustrated in~\citet{TianLewis-BeckNiemiMeeker2022} for some simple
reliability applications, but the ideas there easily extend to more
complicated models.

Computations for our hierarchical POD model were performed in \Rsoftware{}
using the \texttt{rstan} interface to Stan. Based on the specified Stan model
and data, Stan produces a specified number of draws from the joint posterior.
For exploratory analyses, $B=10{,}000$ draws is typically sufficient
when autocorrelation is not too large.
A larger number might be needed to prepare final
results so that all significant digits in reported results are
correct. For the results given here, we used $B=40{,}000$ draws.

Draws from the joint posterior are then used to compute posterior
draws of quantities of interest (e.g., POD for our examples).
For example, POD in~\eqref{equation:unconditional.pod.integral} is a
function of the five population-level parameters $\mu$, $\sigma$,
$\sigma_\tau$, $\sigma_\delta$, and $\sigma_\gamma$. Let the
corresponding posterior draws be $\mu^{[b]}$, $\sigma^{[b]}$,
$\sigma_\tau^{[b]}$, $\sigma_\delta^{[b]}$, and
$\sigma_\gamma^{[b]}$, $b=1,\dots,B$. These can be used to compute
\begin{equation*}
\sigma_{\psi}^{[b]} =
\sqrt{
\left(\sigma_\tau^{[b]}\right)^2
+\left(\sigma_\delta^{[b]}\right)^2
+\left(\sigma_\gamma^{[b]}\right)^2
}, \quad b=1, \dots, B.
\end{equation*}
Then for a given flaw-size value $x$, one can compute
\begin{equation*}
\POD^{[b]}(x)  =
\int_{-\infty}^{\infty}
\Phi\left[
\frac{x-\left(\mu^{[b]} + \sigma_\psi^{[b]} z\right)}{\sigma^{[b]}}
\right]
\phi_{\norm}(z)\,dz, \quad b=1, \dots, B.
\end{equation*}
When  $\epsilon_{ijk} \sim \Normal(0, \sigma)$,
\eqref{equation:probit.unconditional.pod} can be used in a similar
manner, avoiding the numerical integration.

The empirical median of $\POD^{[b]}(x)$, $b=1,\dots,B$, is used as
the point estimate of $\POD(x)$, and the empirical 0.05 quantile of
the $B$ draws is the 95\% lower credible bound for $\POD(x)$. The
computation is repeated for a vector of $x$ values (100 or 200 is
usually sufficient) spanning the range of interest, with the range
chosen so that the lower credible bound for $\POD(x)$ extends from
near zero to near one. The resulting point estimates and lower
credible bounds can then be plotted against $x$. The value of
$a_{90}$ is obtained by interpolating where the point estimate
reaches 0.90, and $a_{90/95}$ is obtained by interpolating where the
95\% lower credible bound reaches 0.90.

Similar computations can be done for the conditional POD for a
particular inspector based on the equations in
Section~\ref{section:conditional.for.inspector.POD}.
For example, for a given inspector $j$ and flaw size $x$,
\eqref{equation:normal.conditional.inspector.pod} is evaluated as
\begin{equation*}
\POD_{j}^{[b]}(x)=
\Phi_{\norm}\left[
\frac{x-(\mu^{[b]}+\delta^{[b]}_{j})}
     {\sqrt{
         \left(\sigma^{[b]}_{\tau} \right)^{2} +
         \left(\sigma^{[b]}_{\gamma} \right)^{2} +
         \left(\sigma^{[b]} \right)^{2}
       }
     }
     \right], \quad b=1, \dots, B.
\end{equation*}

\section{Experimental Study and Statistical Estimation}
\label{sec:experimental-study}

This section applies the statistical methods described in
Section~\ref{section:statistical.models.hit.miss.pod.data} to the
inspection data obtained from X-ray images generated using the
procedures described in Section~\ref{sec:methods}. We first describe
the inspection experiment and the resulting hit-miss data. We then
illustrate POD estimation using data from a single inspection and
compare the results with those from an analysis of the complete
repeated-measures data. The repeated-measures model is used to
estimate the population-mean POD and to evaluate the relative
contributions of the different sources of variability. The section
concludes with a discussion of the experimental validation of XRSIM
and the practical implications for model-assisted POD studies.

\subsection{Experimental Setup}
\label{sec:experimental-setup}

The inspection experiment used synthetic X-ray images generated with
XRSIM for weld cracks of varying depth, width, location and
thickness condition. The images were calculated from the simulator
while keeping a record of the actual location of the flaws. A file
name with all key parameters embedded into the file name was used to
map to a set of file names with a number indicating the order to be
viewed. That order was regenerated for each view of the images.

To generate the hit/miss data, we used five inspectors, each viewing
the set of synthetic images twice, approximately one week apart. The
viewing was done at the same computer, ensuring that
the viewing environment was controlled. Two of the five inspectors
in the study were inexperienced and required training to use the
MaxIm DL view software and training in what to look for. Not
surprisingly, the results for the inexperienced inspectors were not
as good as for the more experienced viewers. The control of the LUT
(Section~\ref{sec:image-viewing-protocol}) was crucial and after it
was discovered that allowing individual control resulted in a far
greater number of missed flaws we determined that the test viewing
needed to be redone with the LUT variability controlled.  Results
from the early protocol-development inspections were not included in
the final dataset.

\subsubsection{Hit/miss inspection results}
\label{sec:hit-miss-results}
\begin{figure}[htbp]
\centering
\includegraphics[width=0.92\linewidth]{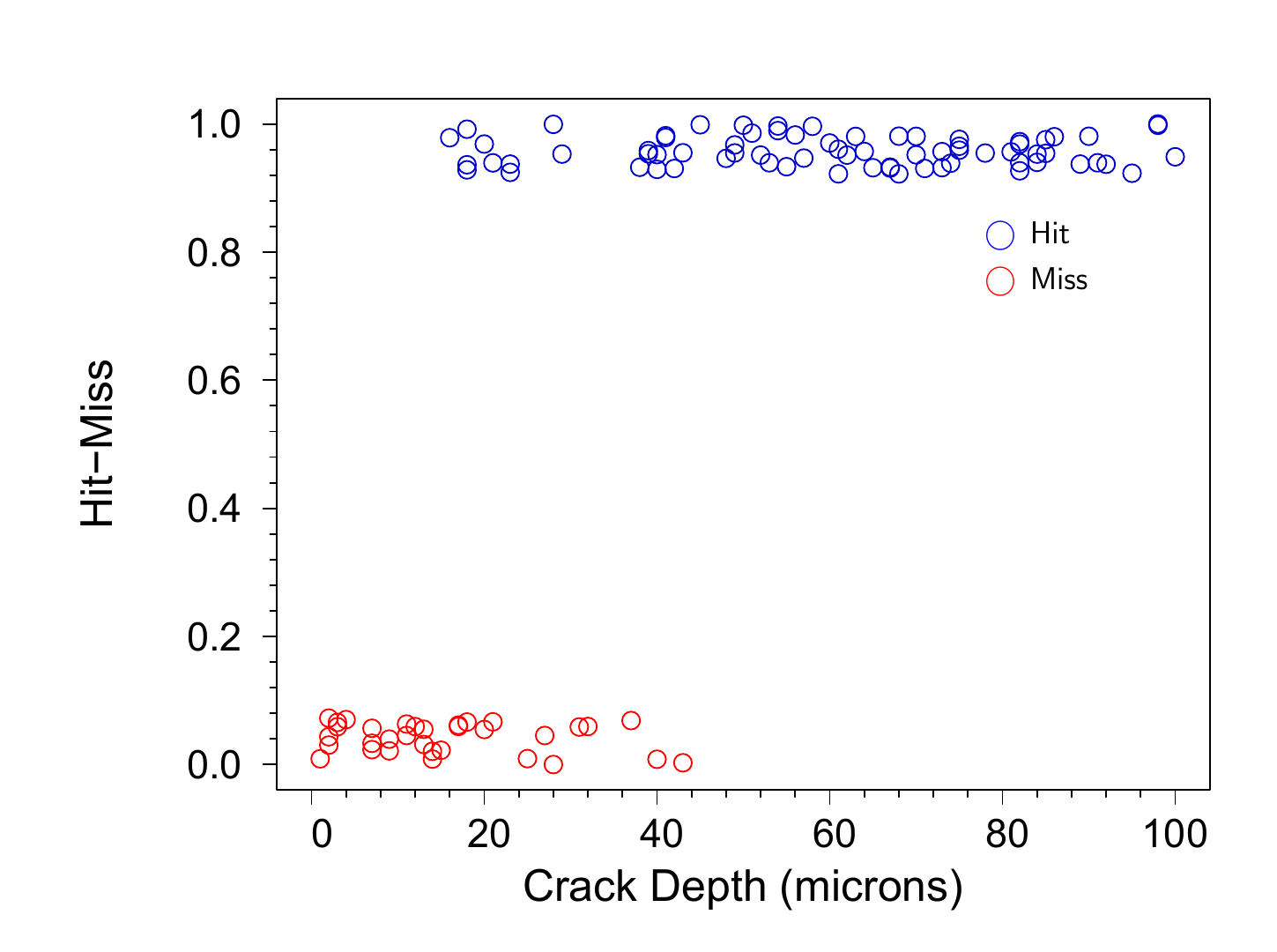}
\caption{Hit-miss data (with jitter) from the first inspection of
  the set of images by Inspector 1. The range of detection from 15 microns
  to 40 microns represents the variation of detectability largely
  due to the weld bead thickness. Welds are typically irregular and
  introduce a thickness variation often of the order of the plates
  being welded. The variation is largely due to the strong
  dependence of the flaw sensitivity to the thickness of the part.}
\label{fig:hit-miss-inspector1}
\end{figure}

The inspection records were collected into hit/miss results where
the tally of the flaws present, found, and missed is recorded. In
Figure~\ref{fig:hit-miss-inspector1}, we see a typical hit/miss plot
of one inspector's inspection of all 105 flaws.
The horizontal axis variable
is the crack depth because that flaw parameter is what controls the
contrast and thus the detectability limit.

\subsubsection{Qualitative observations from the hit/miss inspection results}
\label{sec:results-and-analysis}

There is a wide range of results in flaws missed and detected as a
result of the position in the weld bead. Tight flaws were harder to
see at any thickness. Shallower cracks in thinner areas are easily
seen where the same flaw in thicker areas will not be seen due to
the lack of contrast. We should note that cracks at the weld bead
root were easier to detect than those on the irregular weld
bead. This is a well-known observation that sample irregularity
provides a sort of camouflage effect making the smallest flaws more
difficult to detect.

\subsection{POD Estimates Based on the First Inspection by Inspector~1}
\label{sec:hit-miss-results.op1}
\begin{figure}[htbp]
\centering
\includegraphics[width=0.92\linewidth]{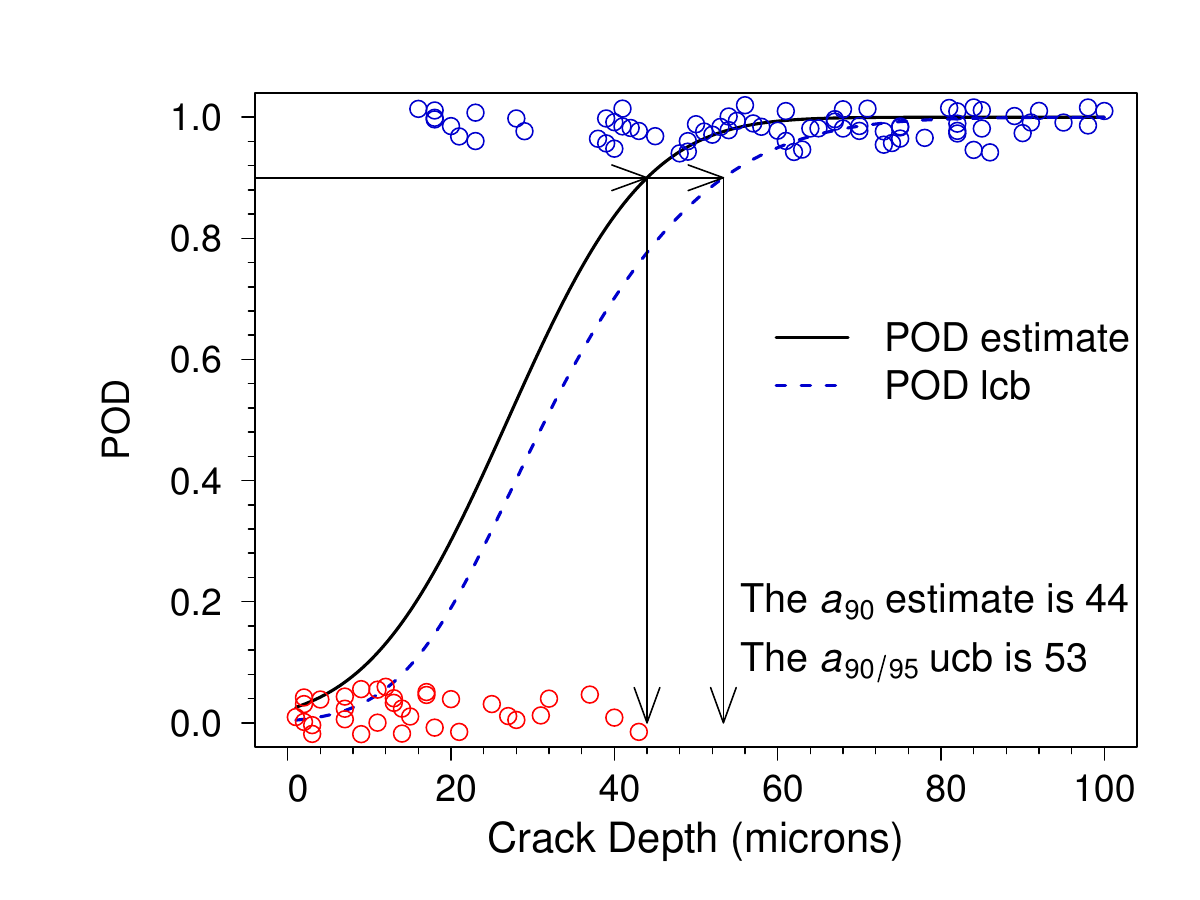}
\caption{Probability of detection estimate, 95\% pointwise lower
  confidence bound (lcb), and corresponding upper confidence bound
  (ucb) for $a_{90}$, denoted by $a_{90/95}$, based on maximum
  likelihood estimation and the Wald approximation for the hit-miss
  data (with jitter) from Inspector~1's first inspection.}
\label{fig:pod-single-inspection}
\end{figure}

This section illustrates the methods in
Section~\ref{section:single-inspection-pod} where each flaw is
inspected once.
Figure~\ref{fig:pod-single-inspection}, which is based on the same
Inspector 1, first inspection
data shown in Figure~\ref{fig:hit-miss-inspector1} (the vertical
positions of the points differ due to different random jitter results),
shows the ML estimate of the POD based on a probit binary
regression model (corresponding to a normal distribution for the
latent detectable flaw-size variable) and the
corresponding $a_{90}$ value. 
Also shown is the
set of lower 95\% pointwise confidence bounds and the corresponding
$a_{90/95}$ value.

\subsection{POD Estimation from Repeated-Measures Data}
\label{sec:repeated-measures-results}
This section illustrates the statistical methods in
Section~\ref{section:repeated-measures-pod} using the complete data
in which all five inspectors inspected each of the 105 flaws
twice. Exploratory analysis of subsets of the data (similar to that
described in Section~\ref{sec:hit-miss-results.op1}) indicated that
the best model for the latent detectable flaw size was a normal
distribution, corresponding to a probit link in the traditional
binary regression model. We use the normal distribution model in the
rest of this section.

\subsubsection{Prior specification}
\label{section:prior.specification}
The Bayesian statistical model requires specification of prior
distributions for the population model parameters. We used weakly
informative priors that place most of the probability on plausible
values for this application while still allowing the data to play
the primary role in determining the posterior distribution over a
wide range. These priors are summarized in
Table~\ref{table:Xray.model.parameter.prior.summary}.
\begin{table}[htbp]
\centering
\begin{threeparttable}
  \caption{Summary of the weakly informative priors for the
    population model parameters.}
\label{table:Xray.model.parameter.prior.summary}
\begin{tabular}{llll}
\toprule
Parameter & Specification type & Distribution & Characteristics \\
\midrule
$\mu$
  & range, dof
  & location-scale $t$
  & range $[1, 100]$; dof = 3 \\
$\sigma$
  & range, dof
  & log-location-scale $t$
  & range $[0.1, 10]$; dof = 3 \\
$\sigma_{\tau}$ (flaw)
  & parameters
  & half-$t$
  & scale = 100; dof = 3 \\
$\sigma_{\delta}$ (inspector)
  & parameters
  & half-$t$
  & scale = 50; dof = 3 \\
$\sigma_{\gamma}$ (flaw/inspector)
  & parameters
  & half-$t$
  & scale = 10; dof = 3 \\
\bottomrule
\end{tabular}
\begin{tablenotes}[flushleft]
\footnotesize
\item \textit{Note:} ``dof'' denotes degrees of freedom.
\end{tablenotes}
\end{threeparttable}
\end{table}
Recall that all of the population model parameters have the same
units as the latent crack depth level that would just be detected
(i.e., microns) and corresponding interpretations in those
units. For $\mu$ and $\sigma$ we follow the approach suggested
in~\citet{TianLewis-BeckNiemiMeeker2022} and specify prior
parameters in terms of a 99\% range defined by the 0.005 and 0.995
quantiles of the prior distribution. This approach makes it
convenient to specify broad prior ranges directly in physically
meaningful units. The ranges in
Table~\ref{table:Xray.model.parameter.prior.summary}
were chosen to include plausible values while putting low
probability on extreme values that would be difficult to reconcile
with the application.

For the random-effect standard deviations, we
use a modification of the suggestion in~\citet{Gelman2006} and
specify half-$t$ distributions with three degrees of freedom.
As a general rule of thumb, a half-$t$ scale parameter that is ten
times larger than the point estimate of the  standard deviation
results in a prior that is approximately flat, relative to the
corresponding marginal posterior.
These
priors place substantial probability near zero while their
relatively heavy tails allow appreciable flaw-to-flaw,
inspector-to-inspector, or flaw/inspector-combination variability
when supported by the data. We chose the scale parameters for the
half-$t$ distributions to reflect our prior knowledge that
flaw-to-flaw variability would be larger than the other sources
of variability, using a scale of 100 for $\sigma_{\tau}$,
50 for $\sigma_{\delta}$, and 10 for $\sigma_{\gamma}$.

\subsubsection{Posterior distribution summary}
\label{section:posterior.distribution.summary}
\begin{figure}[htbp]
\centering
\includegraphics[width=0.92\linewidth]{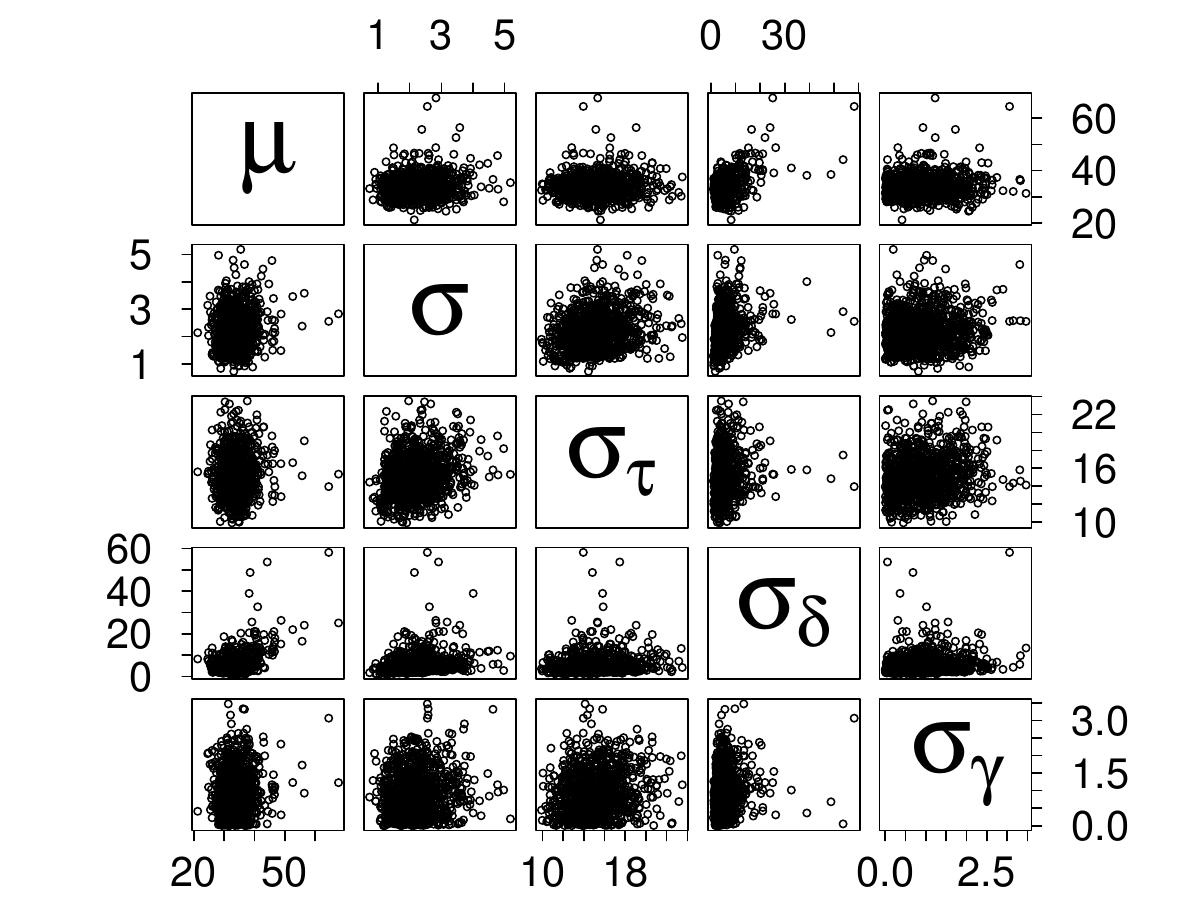}
\caption{Pairs plot of the first 1,000 posterior draws for the
  population model parameters.}
\label{fig:repeated.pairs}
\end{figure}

We used Stan \citep{Stan-software} to generate posterior
draws. After 10,000 warmup iterations, we retained every fifth
post-warmup draw to reduce autocorrelation among the retained draws
and thereby reduce the number of draws that needed to be stored and
processed, resulting in 20,000 draws from the joint posterior
distribution. Traceplots of the posterior draws and other
MCMC diagnostics (not shown here) indicated that the posterior draws
were mixing well.  Figure~\ref{fig:repeated.pairs} is a pairs plot
of the first 1,000 posterior draws for each parameter. The plot
shows little pairwise posterior correlation among these parameters.

Table~\ref{table:Xray.model.parameter.estimation.summary}
summarizes the posterior distribution.

\begin{table}[htbp]
  \caption{Bayesian estimates and credible intervals for the
    population model parameters.}
\label{table:Xray.model.parameter.estimation.summary}
\centering
\begin{tabular}{lrrrr}
\toprule
& & \multicolumn{2}{c}{95\% Credible interval} & \\
Parameter & Estimate (median) & Lower & Upper & $n_{\mathrm{eff}}$ \\
\midrule
$\mu$
  & 33.41
  & 26.95
  & 42.47
  & 30004 \\
$\sigma$
  & 2.09
  & 1.17
  & 3.57
  & 35281 \\
$\sigma_{\tau}$ (flaw)
  & 14.85
  & 11.34
  & 20.23
  & 33961 \\
$\sigma_{\delta}$ (inspector)
  & 4.12
  & 1.64
  & 16.05
  & 32367 \\
$\sigma_{\gamma}$ (flaw-inspector)
  & 0.78
  & 0.04
  & 2.37
  & 35496 \\
\bottomrule
\end{tabular}
\end{table}
The $\muhat=33.41$ microns (median of the critical flaw depth) and
 $\sigmahat=2.09$ microns (controlling how fast the POD curve increases)
 parameters describe a baseline POD when all of the random effects
 are fixed at zero. As expected, $\sigmahat_{\tau}=14.85$ microns
 (flaw-to-flaw variability) dominates the other sources of
 variability. The $\sigmahat_{\delta}=4.12$ microns estimate
 suggests important inspector variability.  The relatively small
 estimate $\sigmahat_{\gamma}=0.78$ microns suggests less additional
 variability associated with particular flaw-inspector
 combinations.  The $n_{\mathrm{eff}}$ values are a little less than
 $B=40,000$ because of some remaining autocorrelation in the
 posterior draws.

\subsubsection{POD estimation from repeated measures data}
\label{section:POD.estimation.repeated.measures.data}
\begin{figure}[htbp]
\centering
\includegraphics[width=0.92\linewidth]{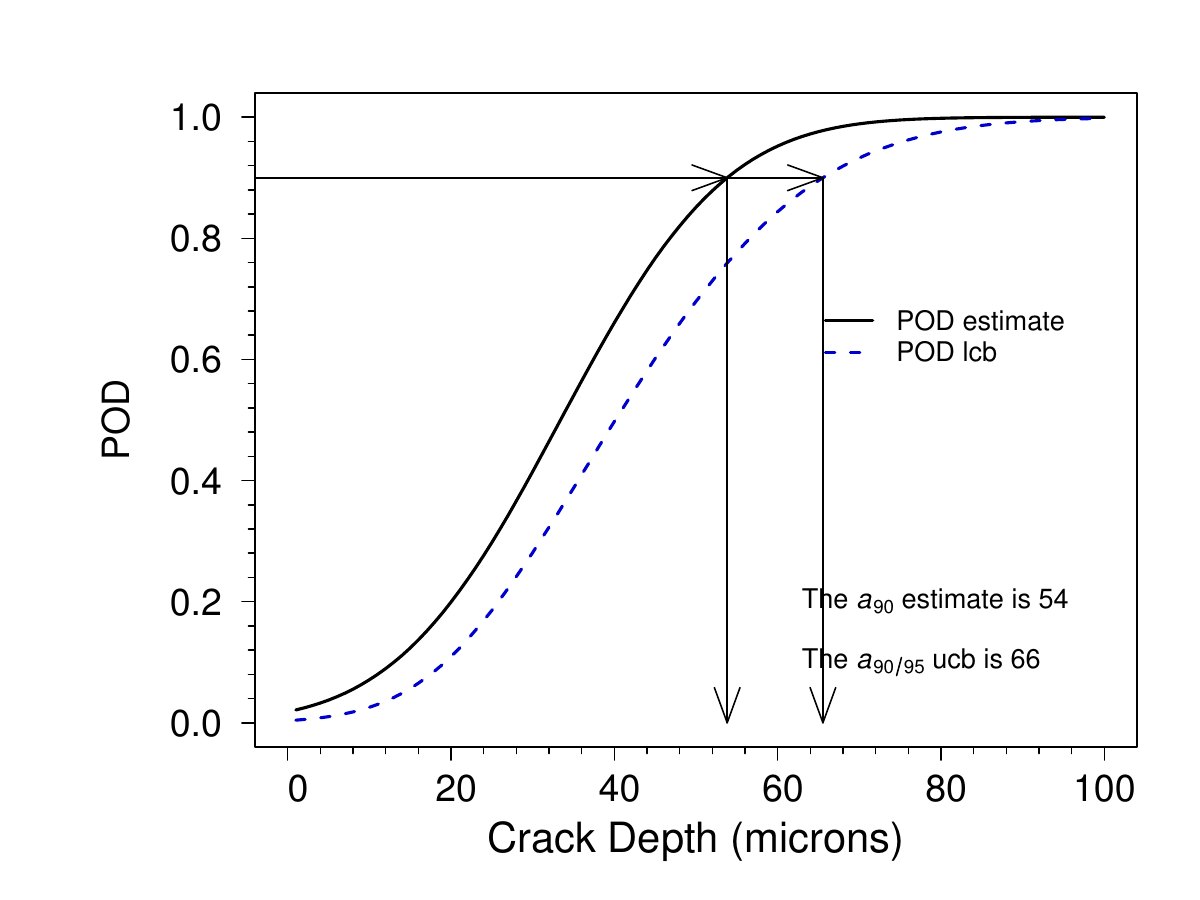}
\caption{Population-mean probability of detection estimate, 95\%
  pointwise lower credible bound (lcb), and corresponding upper
  credible bound (ucb) for $a_{90}$, denoted by $a_{90/95}$, from
  the repeated-measures model for the inspection data.}
\label{fig:pod-repeated}
\end{figure}

We use the posterior draws of the model parameters to compute
posterior draws of the POD function, leading to POD estimates and
lower credible bounds, as described in
Section~\ref{section:statistical.estimation.pod.model}.
Figure~\ref{fig:pod-repeated} shows the POD estimate resulting from
fitting the repeated-measures model to the data.
The plot also illustrates the computation of the estimate of
$a_{90}$ and the corresponding upper credible bound (ucb)
$a_{90/95}$.

\subsubsection{Evaluation of inspector effects and
  inspector-specific POD curves}
\label{sec:repeated.inspector.effects}
\begin{figure}[htbp]
\centering
\includegraphics[width=0.92\linewidth]{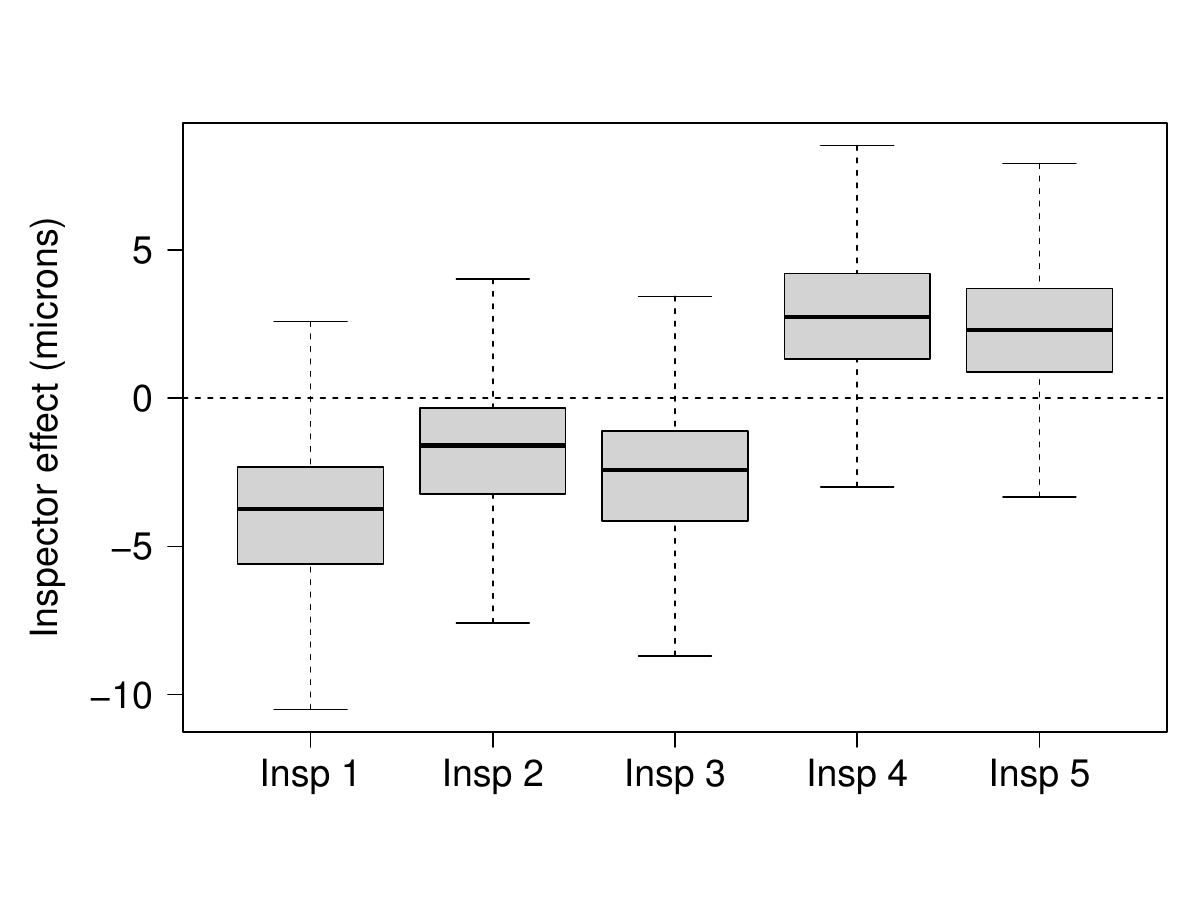}
\caption{Box plots of the posterior draws for the inspector effects
  for each inspector.}
\label{fig:Xray.repeated.inspector.effects}
\end{figure}

Figure~\ref{fig:Xray.repeated.inspector.effects} gives box plots of
the posterior draws for the inspector effects for each inspector.
These plots show appreciable differences among the five inspectors.
Because the effects are expressed on the latent crack-depth scale, a
negative inspector effect corresponds to a lower-than-average
detection threshold and thus better detection performance. A
positive effect corresponds to a higher-than-average detection
threshold and poorer detection performance. The difference between
the median effects for Inspectors~1 and~4, for example, is
approximately 6~microns, indicating a practically important amount
of inspector-to-inspector variability in the crack depth required
for detection. We also computed posterior draws of linear contrasts
between the effect for each inspector and the mean effect for the
other inspectors. In the results (not shown here), the 95\% credible
intervals for those contrasts did not include zero for Inspectors~4
and~5, indicating that their effects differed from the mean effect
of the other inspectors. As noted in
Section~\ref{sec:experimental-setup}, two of the five inspectors in
the study were inexperienced and had limited pre-inspection
training.

\begin{figure}[htbp]
\begin{tabular}{cc}
(a) & (b) \\[-3.2ex]
\includegraphics[width=0.46\linewidth]{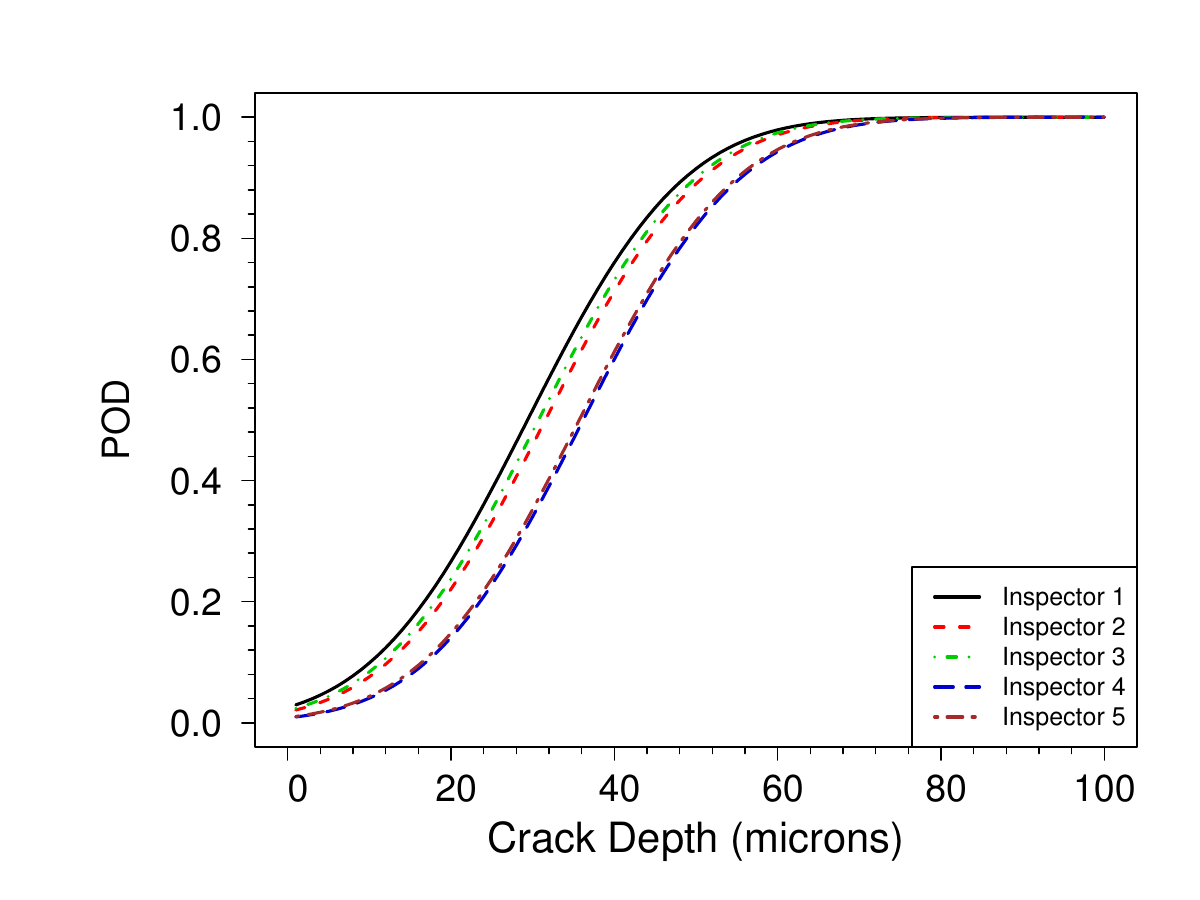} &
\includegraphics[width=0.46\linewidth]{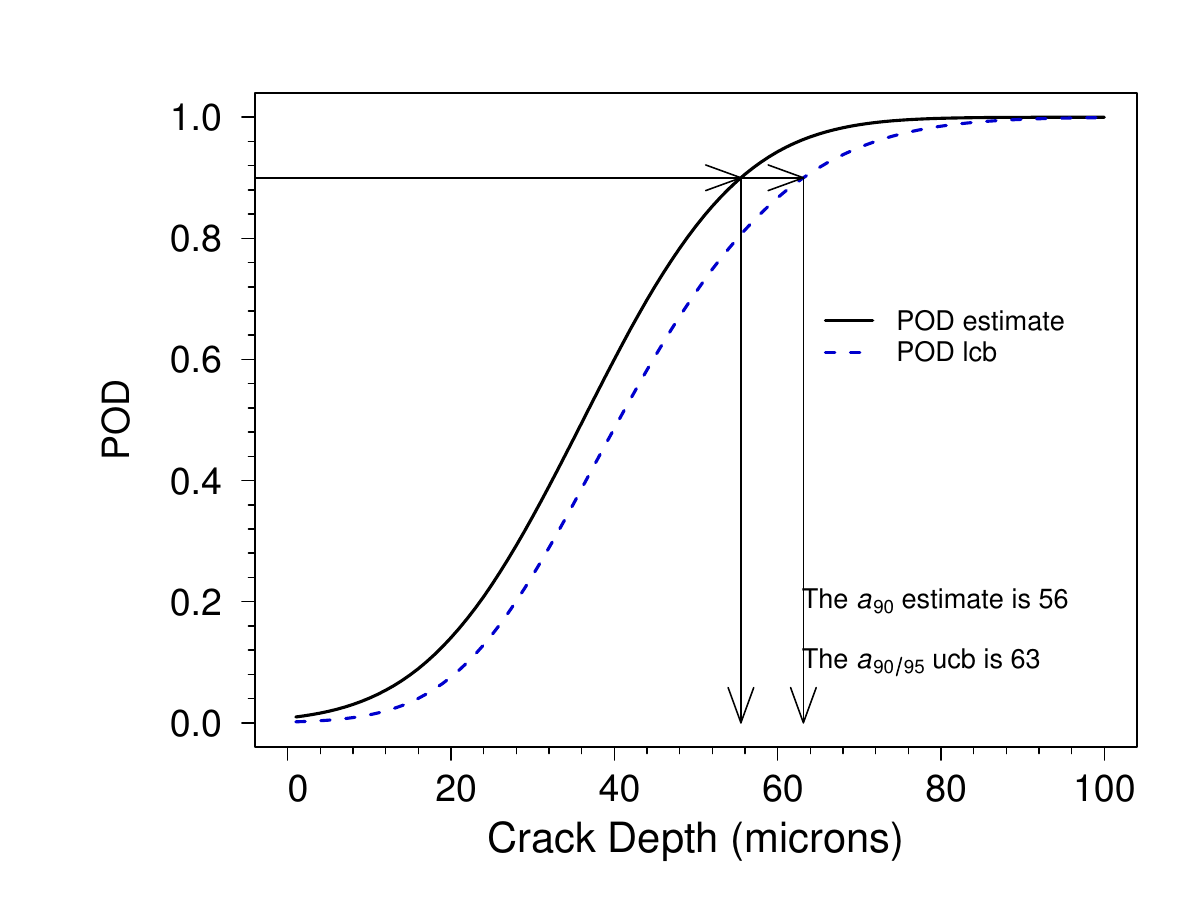}
\end{tabular}
\caption{Probability of detection (POD) estimates for all five
  inspectors~(a), and the POD estimate, 95\% pointwise lower credible bound (lcb), and
  corresponding upper credible bound (ucb) for $a_{90}$, denoted by
  $a_{90/95}$, for Inspector~4~(b).}
\label{figure:Xray.conditional.inspector.POD}
\end{figure}

Figure~\ref{figure:Xray.conditional.inspector.POD}a shows
inspector-specific POD estimates for all five inspectors.
Figure~\ref{figure:Xray.conditional.inspector.POD}b shows estimates
of POD and $a_{90}$, together with the $a_{90/95}$ upper credible
bound, for Inspector~4, the inspector having the poorest estimated
performance in the experiment. Thus, among the five inspectors,
Inspector~4 provides the most conservative inspector-specific POD
curve.

\subsection{Model Validation and Practical Implications}
\label{sec:model-validation}
\begin{figure}[htbp]
\centering
\includegraphics[width=0.92\linewidth]{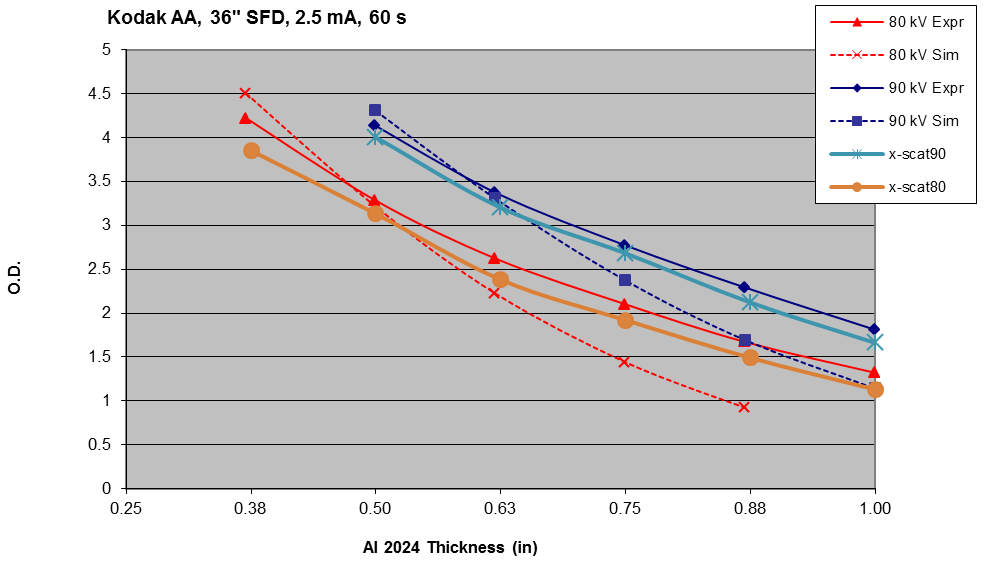}
\caption{Experimental validation of the X-ray radiographic
  simulation program is shown in a validation study for an
  industrial inspection. The solid lines show simulated and
  experimental results for film density over a range of aluminum
  thickness with all of the key physics variables modeled.
  The dotted lines show the simulation result without using a model for scattering.}
\label{fig:xrsim-validation}
\end{figure}

\begin{figure}[htbp]
\centering
\includegraphics[width=0.92\linewidth,height=0.68\textheight,keepaspectratio]{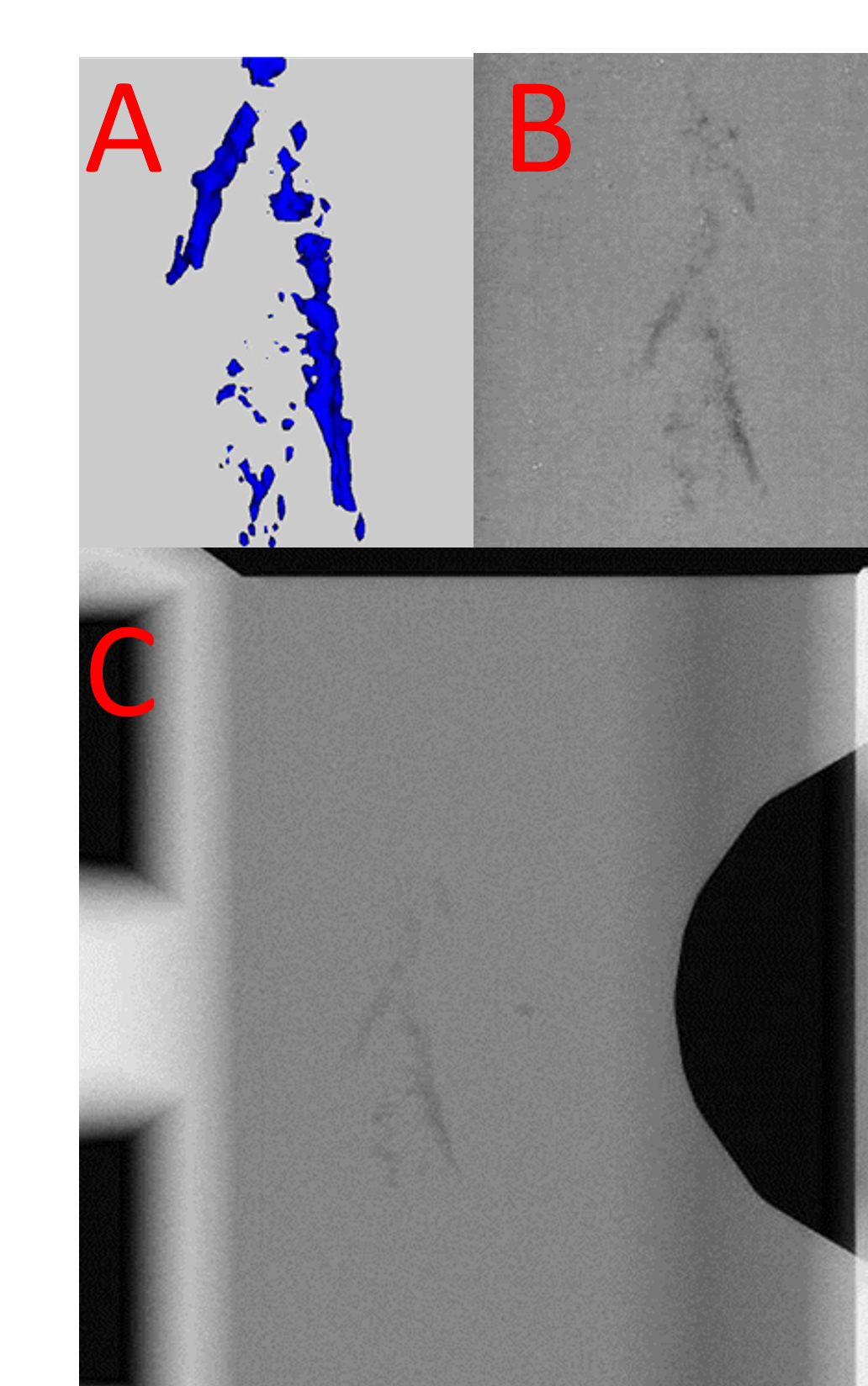}
\caption{Demonstration of the ability to represent complex and
  realistic flaw geometry. Panel~A shows a computer-aided design
  (CAD) model of a casting porosity flaw obtained by reverse
  engineering computed tomography data. Panel~B shows the
  corresponding experimental radiograph, and Panel~C shows the
  radiograph simulated using XRSIM.}
\label{fig:complex-flaw-geometry}
\end{figure}

A key requirement for the use of XRSIM is experimental validation
of the simulation model. Figures~\ref{fig:xrsim-validation} and
\ref{fig:complex-flaw-geometry} show two examples of such
validation.

Figure~\ref{fig:xrsim-validation} shows a comparison
that is representative of the agreement observed between XRSIM
simulation results and experimental measurements.

The solid lines show agreement between the experimental measurements
and XRSIM predictions to within approximately 2\%. The dotted lines show
XRSIM results obtained without a scattering model,
illustrating two points: first, the need for experimental validation
and second, the quantitative effect of not modeling scattering.
It should be noted that the magnitude of the scattering
shown in Figure~\ref{fig:xrsim-validation} is
small. In some inspection conditions involving thick samples
and high-atomic-number materials, such as steel or nickel,
scattered photons can account for approximately 50\% of the
X-ray photons reaching the detector.

Experimental validation quantifies the discrepancy between
model predictions and experimental measurements. Thus, for the
situation shown in Figure~\ref{fig:xrsim-validation}, omitting
scattering from the model has a relatively small effect on the
predicted detector response.
What is not shown is the effect that scattering has on contrast
sensitivity. We noted that a typical target for detector response
is roughly 80\% of the saturation signal. In the X-ray process,
two types of X-ray photons reach the detector: directly transmitted
photons, whose intensity is modulated by attenuation through the
part, and scattered photons that contribute to the background signal.
When the total detector response reaches 80\% of saturation,
the directly transmitted component can be only a fraction of the
total intensity---in severe cases less than half. Thus, although
the total detector response is near the desired level, the
directly transmitted component that carries the flaw contrast
is relatively weak. The scattered photons also contribute noise,
reducing the CNR of the flaw. Consequently, a large scattering
component degrades the ability of the inspection to detect
smaller flaws.

The important consideration in using a simulation model
is to understand which physical processes are represented adequately
and to have experimental data establishing the range of conditions
over which the model provides acceptable agreement.
The ability of XRSIM to turn off scattering effects produced the
dotted lines in Figure~\ref{fig:xrsim-validation}. This capability
allows for a rapid evaluation of the magnitude of the scattering
effects, providing a useful tool for determining the inspection
protocol.

The second validation example is shown in
Figure~\ref{fig:complex-flaw-geometry} and demonstrates the ability
of XRSIM to represent complicated flaw morphology.
Panel~A shows a CAD model obtained from a 3D computed
tomography scan of a casting porosity flaw. The computed
tomography data were converted to a 3D CAD model suitable for import
into XRSIM. Comparison of the experimental radiograph in Panel~B
with the simulated radiograph in Panel~C shows that the complex flaw
morphology is reproduced well in the simulated radiograph.
This capability is important when POD estimates are used in reliability
calculations. Few real flaws have simple geometries.
Idealized fabricated flaws such as uniform notches or flat-bottom holes
are useful for demonstrating inspection-system capability,
but do not capture the effect of realistic flaw morphology on detectability.
The results in Figure~\ref{fig:complex-flaw-geometry}
demonstrate the potential of simulation tools to help distinguish
poor detectability caused by flaw characteristics or inspection
conditions from misses attributable to human inspectors.
The importance of crack tightness in this study illustrates this point.

\section{Discussion, Conclusions, and Areas for Future Research}
\label{sec:discussion}

The use of NDE simulation programs such as XRSIM relies on adequately
capturing the key physics involved in image formation. Such simulation
programs must adequately represent the output of typical equipment
used in NDE inspections as well as important part characteristics,
including part material and the morphology of typical flaws. In an
experimental determination of POD, the goal is to design an
experiment that captures the important sources of variability
affecting image formation as well as the human variability associated
with an inspector's ability to recognize a feature in an image as a
flaw.

XRSIM can represent many of the physical sources of variability in
an inspection. For example, changes in part orientation relative to
the X-ray beam, an important factor in detecting cracks, can readily
be incorporated. Flaw characteristics and flaw location can also be
varied when generating the images. Detector characteristics,
especially noise, are important in determining the detection limit
for a given flaw. An important advantage of XRSIM is its ability to
model the image-formation process in considerably more detail than
is economically feasible in many experimental POD studies. For
example, the cracks used in this study were chosen to mimic the
relatively simple shapes of manufactured flaws commonly used in
experimental POD studies. It is well known, however, that such
idealized shapes are not representative of the irregular flaws that
can result from aging and other degradation processes. In
experimental POD studies, fabricating hundreds of flaws spanning
wide ranges of shape, orientation, and location within a part
would be prohibitively expensive. Such variation can be captured
much more readily using tools such as XRSIM.

XRSIM itself does not model the human decision process or the
associated inspector-to-inspector and inspection-to-inspection
variability. Considerable effort has been devoted in the
medical-imaging community to characterizing observer performance and
radiologist variability
\citep[e.g.,][]{fryback1991,Krupinski2010current,he2013}.
Related work in color science led to the development and subsequent
refinement of standard human-observer models
\citep{berns2000}. In addition, the present analysis uses crack depth
as the primary POD explanatory variable even though inspection
response can depend on several flaw and image characteristics.

An important opportunity, therefore, is to use XRSIM to generate
realistic images sampled from a suitably defined population of
inspection conditions and then use those images to characterize
human-inspector variability directly. Such studies could ultimately
provide the information needed to develop useful models of a
standard human observer for NDE applications. The same simulation
capability also provides important opportunities for inspector
training. Inspectors could be exposed to realistic examples of the
types of flaws that can occur in particular parts, and experienced
inspectors could periodically refresh their familiarity with
uncommon flaw configurations. New inspectors could similarly be
trained using realistic combinations of parts, flaws, and inspection
conditions rather than relying only on more abstract training
materials.

As described in Section~\ref{sec:paper-organization}, this paper
presents a model-assisted POD (MAPOD) framework based on XRSIM for
image-based human inspection of X-ray weld images.  The study shows
how simulated images can reduce the need to fabricate large numbers
of physical flaw specimens while still allowing important sources of
inspection variability to be represented. For repeated-measures
inspection data, the Bayesian random-effects model used here
accounts for flaw-to-flaw, inspector-to-inspector, and
flaw/inspector-combination variability as well as
inspection-to-inspection variability. Treating repeated inspections
as independent would overstate the amount of information in the data
and result in statistical intervals that are too narrow. The
statistical framework used in this paper provides estimates of the
population-mean POD and also provides a foundation for more conservative
inspection-capability assessments, including inspector-specific and
quantile-based POD measures.

The material in this paper suggests several areas for future
research.
\begin{itemize}
\item
\citet{li2014} introduced the concept of ``quantile POD'' and
applied it to an application involving $\hat{a}$ versus $a$ POD
data. \citet{koh2014,KohMeeker2019} applied similar methods to
hit-miss POD data. The main idea is that, because of the various
sources of variability described in this paper, POD at a given
flaw size varies across the population of flaw characteristics
and inspection conditions.
Traditional POD estimation averages over these sources of
variability to obtain a population-mean POD. Similar to common
practice in product reliability applications
\citep[e.g.,][]{MeekerEscobarPascual2022}, \citet{li2014} argued
that, when safety is involved, it may be more appropriate to report
estimates and lower confidence bounds for a lower quantile of the
distribution of POD. Such quantile POD estimates could be computed
using an extension of the model developed in this paper.

\item
In many applications, machine-learning algorithms are used for
initial screening of images to identify ``regions of interest'' that
are then examined carefully by a human inspector who makes the final
decision. The simulation-based methods described in this paper could
also be used to generate data for training and evaluating the
inspection performance of both the machine-learning algorithm and
the human inspectors involved in the inspection process.

\item
Traditionally, POD has been estimated as a function of one key flaw
characteristic, such as crack length or depth. POD is often an
important input to risk-assessment models, and such models may need
to account for the increased risk associated with other flaw
characteristics, such as orientation and position within a part.  In
principle, POD studies can be extended to model POD as a function of
multiple explanatory variables. Doing this with traditional physical
experiments could be prohibitively expensive.  Conducting such
experiments using simulated flaws in images would be feasible for
X-ray and other image-based inspection methods.  The statistical
methods developed in this paper could be extended to analyze such
multivariable POD data.
\end{itemize}

\section*{Acknowledgments}
Part of the research in this paper was supported by the Federal
Aviation Administration under contract number DTFACT-09-C-00006 and
performed at Iowa State University's Center for Nondestructive
Evaluation.  We would like to thank Necip Doganaksoy (Siena
University) and Luis Escobar (Louisiana State University) for
helpful comments on an earlier version of this paper.

\section*{Declaration of Conflicting Interests}
The authors declare that there are no conflicts of interest.

\begin{samepage}
\section*{Data Availability Statement}
The datasets used in the current study are available in the
supplementary materials associated with this article.
\end{samepage}

\bibliographystyle{chicago}
\bibliography{references}

\end{document}